\documentclass[a4paper,11pt]{article}
\pdfoutput=1 

\usepackage{jheppub} 

\usepackage[T1]{fontenc} 

\usepackage{amssymb}
\usepackage{enumerate} 

\usepackage{multirow} 
\usepackage{booktabs} 
\usepackage{makecell} 

\usepackage{amsmath} 
\usepackage{xspace} 
\usepackage{tensor} 

\def\eqspace{\quad\quad\quad}
\def\LOone{${\rm LO}_1$\xspace}
\def\LOtwo{${\rm LO}_2$\xspace}
\def\LOthree{${\rm LO}_3$\xspace}
\def\LOfull{${\rm LO}$\xspace}
\def\NLOone{${\rm NLO}_1$\xspace}
\def\NLOtwo{${\rm NLO}_2$\xspace}
\def\NLOthree{${\rm NLO}_3$\xspace}
\def\NLOfour{${\rm NLO}_4$\xspace}
\def\NLOqcd{${\rm NLO}_{\rm QCD}$\xspace}
\def\NLOfull{${\rm NLO}$\xspace}
\def\NLOprd{${\rm NLO}_{\rm prd}$\xspace}
\def\helacdipoles{\textsc{Helac-Dipoles}\xspace}
\def\Recola{\textsc{Recola}\xspace}

\title{Complete NLO corrections to top-quark pair production with isolated photons}

\author{Daniel Stremmer}
\author{and Malgorzata  Worek}
\affiliation{ Institute for Theoretical Particle Physics
and Cosmology, RWTH Aachen University, \\D-52056 Aachen, Germany}

\emailAdd{daniel.stremmer@rwth-aachen.de}
\emailAdd{worek@physik.rwth-aachen.de}

\abstract{
We compute for the first time the so-called complete NLO corrections to top-quark pair production with one and two isolated photons  in the di-lepton top-quark decay channel. The Narrow Width Approximation is used for the modeling of unstable top quarks and $W$ bosons. Higher-order QCD and EW effects as well as photon bremsstrahlung are consistently included at all stages: in production and top-quark decays. We present results at the integrated and differential fiducial cross-section level for both processes for the LHC Run II center-of-mass energy of $\sqrt{s}=13$ TeV. In addition, we investigate the scale choice in photonic observables. Finally, the individual size of each subleading contribution is discussed in detail and the origin of the main subleading corrections is scrutinised. For the latter case, alternative calculations are performed in which the subleading NLO corrections are  included only in the production of $t\bar{t}\gamma$ and $t\bar{t}\gamma\gamma$.}

\dedicated{\rm TTK-24-06, P3H-24-005}


\begin{document} 
\maketitle
\flushbottom

%
\section{Introduction}
\label{sec:introduction}
%

Of all the associated production of a top-quark pair with one electroweak gauge boson $(pp \to t\bar{t}V, \, V=Z,W^\pm, \gamma)$ at the LHC, the $pp\to t\bar{t}\gamma$ process has the largest cross section \cite{Maltoni:2015ena,Schwienhorst:2022yqu} and is the most interesting to model. First of all, already at the  lowest order in  the perturbative expansion, the $pp\to t\bar{t}\gamma$ process requires various cuts to keep the prompt photon well isolated and render the whole process infrared (IR) finite, even in the presence of stable top quarks. Secondly, due to its massless nature, the photon can be emitted from all stages of the process, i.e. in the production of $t\bar{t}$ and top-quark decays. In the latter case, prompt photons can originate not only from top quarks, but also from their decay products, including the bottom quarks, $W^\pm$ gauge bosons and charged leptons. In addition, they can be radiated from incoming partons. It is a well-known fact that a large fraction of isolated photons comes from radiative decays of top quarks \cite{Melnikov:2011ta,Bevilacqua:2019quz}. With fairly inclusive cuts applied on the finale states, that are currently used in measurements of inclusive and differential cross sections of $t\bar{t}\gamma$ production, the contribution of photons at the decay stage reaches almost $50\%$. Consequently, proper modeling of $pp\to t\bar{t}\gamma$ is very challenging. The $pp\to t\bar{t}\gamma$ process probes the $t-\gamma$ electroweak coupling and provides a direct way to measure the top-quark electric charge \cite{Baur:2001si}. The latter is known to be consistent with the Standard Model (SM), although it was measured only indirectly in the production of $t\bar{t}$ \cite{D0:2014oid,ATLAS:2013mkl}. Precise measurements of the $t-\gamma$ coupling at the LHC serve as an additional test of the SM. However, any deviations from the SM prediction, for example in the $p_T$ spectrum of the photon, could point to new physics through anomalous dipole moments of the top quark \cite{Baur:2004uw,Bouzas:2012av,Rontsch:2015una,
Schulze:2016qas,BessidskaiaBylund:2016jvp}. In addition, the $pp\to t\bar{t}\gamma$ process plays an important role in studies of the top-quark charge asymmetry $(A_C^t)$ \cite{Aguilar-Saavedra:2014vta,Aguilar-Saavedra:2018gfv,Bergner:2018lgm,Pagani:2021iwa}. Indeed, it provides complementary information to the measured asymmetries in $t\bar{t}$ production, where $A_C^t$ appears for the first time at next-to-leading order (NLO) only. Contrary, for the $pp \to t\bar{t}\gamma$ process this asymmetry is present already at LO in quark-induced subprocesses due to the interference effects between Feynman diagrams in which the photon is emitted from quarks in the initial state and diagrams in which it is emitted from quarks in the final state. The overall asymmetry in the $pp \to t\bar{t}\gamma$ process at the LHC with $\sqrt{s} = 13$ TeV has a negative value and is of the order of $1\%-2\%$ depending on the fiducial phase-space regions that are scrutinised \cite{Bergner:2018lgm,Pagani:2021iwa}. However, $A_C^t$ can be modified by beyond the SM (BSM) physics. Indeed, substantial deviations from the SM prediction can be expected in the case of BSM models with a light colour octet \cite{Barcelo:2011vk,MarquesTavares:2011spp,Alvarez:2011hi,Aguilar-Saavedra:2011lzv} or an additional $Z^\prime$ \cite{Alvarez:2013jqa}. In both BSM cases, the absolute value of $A_C^t$ is predicted to be much smaller \cite{Aguilar-Saavedra:2014vta}. 

In the case of the $pp \to t\bar{t} \gamma\gamma$ process, the situation is only somewhat similar to the case of $t\bar{t}\gamma$. Among all the associated production of a top-quark pair with two gauge vector bosons, i.e. $t\bar{t}W^+W^-$, $t\bar{t}ZZ$, $t\bar{t}\gamma\gamma$, $t\bar{t}W^\pm \gamma$, $t\bar{t}W^\pm Z$ and $t\bar{t}Z\gamma$, the  $pp \to t\bar{t}\gamma\gamma$ process is the second largest \cite{Maltoni:2015ena}. The importance of  $pp \to t\bar{t} \gamma\gamma$ stems from the fact that it is the main (irreducible) background process in SM Higgs-boson studies for  the $pp \to  t\bar{t}H$ signal process in the $H\to \gamma\gamma$ decay channel. However, the similarities to the production of $t\bar{t}\gamma$ are due to the distribution of photons in the $pp \to t\bar{t}\gamma\gamma$ process and the importance of the contribution, which is not entirely related to the production stage of the process. The effect of photon bremsstrahlung in the $pp \to t\bar{t} \gamma\gamma$ process has recently been studied in detail \cite{Stremmer:2023kcd}. It has been shown  that the so-called  mixed contribution, in which two photons occur simultaneously in the production and decay of the $t\bar{t}$ pair, is the dominant contribution at the integrated and differential fiducial cross-section  level.

Evidence for the production of a top-quark pair in association with an isolated photon has already been reported in $p\bar{p}$ collisions at the Tevatron collider at $\sqrt{s}=1.96$ TeV by the CDF Collaboration \cite{CDF:2011ccg}. The production of $t\bar{t}\gamma$ final states  has been observed  for the first time at the LHC at $\sqrt{s}=7$ TeV by the ATLAS Collaboration \cite{ATLAS:2015jos}. To date, both ATLAS and CMS have observed the production of $t\bar{t}\gamma$  at the LHC at $\sqrt{s}=8$ TeV  and $\sqrt{s}=13$ TeV \cite{ATLAS:2017yax,CMS:2017tzb,CMS:2021klw,ATLAS:2018sos,ATLAS:2020yrp,CMS:2022lmh}, respectively. So far, no significant deviations from the SM predictions have been found, even though the measured cross sections are larger than theoretical predictions. In addition, within the current and still rather large uncertainties, the analysed differential cross-section distributions have also been fairly well described by the NLO theory predictions. The measurements in the $pp\to t\bar{t}\gamma$ process have also been interpreted in the framework of the standard model effective field theory, where rather stringent limits on the two relevant Wilson coefficients have been found. Furthermore, the measurement of the top-quark charge asymmetry in $pp\to t\bar{t}\gamma$ has recently been performed by the ATLAS collaboration \cite{ATLAS:2022wec}. Also in this case  the measurement is compatible with the SM predictions within the present uncertainties. However, in all these cases, current precision is still limited, mainly due to statistical and various systematic uncertainties, leaving room for potential future improvements. Finally, the $pp \to t\bar{t} \gamma\gamma$ process has not yet been experimentally observed at the LHC.

On the theory side, NLO QCD corrections to the $pp\to t\bar{t}\gamma$ production process with stable top quarks have been calculated over ten years ago \cite{Duan:2009kcr,Duan:2011zxb} and calculated afresh in Ref. \cite{Maltoni:2015ena}. The results with  NLO electroweak (EW) corrections have also been delivered later \cite{Duan:2016qlc}. The results presented there have shown that for differential cross-section distributions the NLO EW corrections are significant in the high energy region due to the EW Sudakov effect. Recently, the so-called complete NLO predictions for $pp\to t\bar{t}\gamma$ have been calculated \cite{Pagani:2021iwa}, again for stable top quarks only. In addition to the NLO QCD and EW corrections at ${\cal O}(\alpha_s^3\alpha)$ and ${\cal O}(\alpha_s^2\alpha^2)$ respectively, various subleading contributions, along with their higher-order effects, have been taken into account to form the complete NLO result. In detail, the set of all the possible contributions at ${\cal O}({\alpha_s^2\alpha})$, ${\cal O}({\alpha_s\alpha^2})$ and ${\cal O}({\alpha_s^0\alpha^3})$ at LO,  where the middle one results from the interference of the other two results at the amplitude level and from the photon initiated partonic process $g\gamma\to t\bar{t}\gamma$, as well as at ${\cal O}({\alpha_s^3\alpha})$, ${\cal O}({\alpha_s^2\alpha^2})$, ${\cal O}({\alpha_s\alpha^3})$ and ${\cal O}({\alpha_s^0\alpha^4})$ at NLO  is what is denoted as the complete NLO result. Finally, the approximate NNLO cross section, with second-order soft-gluon corrections added to the NLO result including QCD and EW corrections, has been calculated in Ref. \cite{Kidonakis:2022qvz}. On top of the stable top-quark approximation, various predictions that take into account top-quark decays are available in the literature. First,  NLO QCD theoretical predictions for $t\bar{t}\gamma$ have been matched with the \textsc{Pythia} parton shower program \cite{Kardos:2014zba}. In this approach top-quark decays have been treated in the parton-shower approximation omitting $t\bar{t}$ spin correlations and photon emission in the parton-shower evolution. More realistic predictions at NLO in QCD have been presented in Ref.~\cite{Melnikov:2011ta}. In this case, top-quark decays in the Narrow Width Approximation (NWA) have been included, maintaining $t\bar{t}$ spin correlations in the top-quark decay products. In addition, photon radiation off charged top-quark decay products has been incorporated. A complete description of  the $pp\to t\bar{t}\gamma$ process in the di-lepton top-quark decay channel at NLO in QCD has also been finally provided \cite{Bevilacqua:2018woc}. This calculation is based on matrix elements for the $e^+\nu_e \mu^- \bar{\nu}_\mu b\bar{b}\gamma$ final state including all resonant and non-resonant Feynman diagrams, interferences and off-shell effects of the top quarks and $W$ gauge bosons. Finally, a dedicated comparison between  the full off-shell calculation and the results in the NWA has also been carried out \cite{Bevilacqua:2019quz}. For the $pp\to t\bar{t}\gamma\gamma$ process the situation is much simpler.
NLO QCD predictions for the process with stable top quarks have already been known for some time and have been further matched to parton shower programs \cite{Alwall:2014hca,Kardos:2014pba,Maltoni:2015ena,vanDeurzen:2015cga}. 
In addition to higher-order QCD effects,  NLO EW corrections for $t\bar{t}\gamma\gamma$ have recently been reported \cite{Pagani:2021iwa}, but again only for stable top quarks.  Finally, very recently NLO QCD corrections in the NWA for the di-lepton and lepton $+$ jet top-quark decay channels have been calculated for the $pp\to t\bar{t}\gamma\gamma$ process  \cite{Stremmer:2023kcd}. 

From the above description, we can quite clearly see a rather diverse picture for $pp\to t\bar{t}\gamma$ and $pp\to t\bar{t}\gamma\gamma$. On the one hand, we already have the complete NLO result for $t\bar{t}\gamma$, albeit only for stable top quarks. On the other hand, no such predictions exist for $t\bar{t}\gamma\gamma$. In both cases, the complete NLO result with more realistic final states  is still missing.  With this article, we want to mitigate the current situation and to calculate for the first time the complete NLO predictions for both $pp\to t\bar{t}\gamma$ and $pp\to t\bar{t}\gamma\gamma$ in the di-lepton top-quark decay channel.  

The aim of our work is, therefore, manifold.  Firstly, we perform a detailed study of the predictions with complete NLO corrections included  for $pp\to t\bar{t}\gamma$ and $pp\to t\bar{t}\gamma\gamma$ in the di-lepton top-quark decay channel. Secondly, as we calculate these higher-order effects within the same framework, while using the same input parameters, we want to examine whether there are common features or significant differences between the theoretical predictions for $pp\to t\bar{t}\gamma$ and $pp\to t\bar{t}\gamma\gamma$. Thirdly, we plan to analyse the impact of subleading NLO corrections on integrated and differential fiducial cross sections and systematically study the theoretical uncertainties due to missing higher orders. Finally, we are going to scrutinise the individual size of each subleading contribution and investigate the origin of the main subleading corrections. For this purpose, we perform alternative calculations for both processes  in which the subleading NLO corrections are included only in the production of $t\bar{t}\gamma(\gamma)$.

The rest of the paper is organised as follows. All leading and subleading contributions at LO and NLO are carefully defined in Section \ref{sec:definition}. In Section \ref{sec:description} we describe the framework of the calculations and discuss various changes as well as  cross-checks that have been performed. All input parameters and fiducial cuts that have been used to simulate detector response are outlined in Section \ref{sec:setup}. Numerical results for the integrated and differential cross sections for the $pp \to t\bar{t}\gamma$ process at the LHC are presented in detail in Section \ref{sec:tta}. Our results for the $pp \to t\bar{t}\gamma\gamma$ process, on the other hand, are discussed in Section \ref{sec:ttaa}. We summarise our findings in Section \ref{sec:sum}.

%
\section{Definition of the leading and subleading contributions}
\label{sec:definition}
%

We calculate the full set of leading and subleading LO contributions and NLO corrections to top-quark pair production with one and two isolated photons at the LHC. We consistently include photon bremsstrahlung as well as QCD and EW corrections in both the production and decay of the top-quark pair. Decays of unstable top quarks and $W$ gauge bosons are treated in the NWA preserving spin correlations, i.e. in the limit when $\Gamma_t/m_t\to 0$  $(\Gamma_W/m_W\to 0)$. In this approximation all contributions without two resonant top quarks (and $W$ gauge bosons) are neglected and the Breit-Wigner propagators lead to delta-functions which force the unstable particles to be on-shell, 
see e.g. \ Ref.~\cite{Denner:1999gp,Denner:2005fg,Melnikov:2009dn,Melnikov:2011qx,Campbell:2012uf,Behring:2019iiv,Bevilacqua:2019quz,Czakon:2020qbd} for more details. We consider the di-lepton top-quark decay channel leading to the following decay chains
\begin{equation}
\begin{split}
pp &\to t\bar{t}(\gamma)\to W^+W^-\,b\bar{b}(\gamma) \to \ell^+\nu_{\ell}\, \ell^-\bar{\nu}_{\ell} \, b\bar{b}\,\gamma +X, \\[0.2cm]
pp &\to t\bar{t}(\gamma\gamma)\to W^+W^-\,b\bar{b}(\gamma\gamma) \to \ell^+\nu_{\ell}\, \ell^-\bar{\nu}_{\ell} \, b\bar{b}\,\gamma\gamma +X, 
\end{split}
\end{equation}
with  $\ell^{\pm}=\mu^{\pm},e^{\pm}$ and where the brackets indicate that  photon bremsstrahlung is allowed at each stage of the process. For brevity, we refer to these processes as $pp \to t\bar{t}\gamma$ and $pp \to t\bar{t}\gamma\gamma$  in the di-lepton top-quark decay channel, respectively. At LO there are in total $5$ ($15$) possibilities or resonance histories from which the photons can be radiated off in the decay chain of $t\bar{t}\gamma(\gamma)$. At NLO, the number of resonance histories increases to $15$ ($35$) for real radiation with an additional photon and to $15$ ($45$) for additional QCD radiation.

%
\subsection{LO contributions}
\label{sec:desc-lo}
%

%
\begin{figure}[t!]
  \begin{center}
    \includegraphics[trim= 20 735 20 10, width=\textwidth]{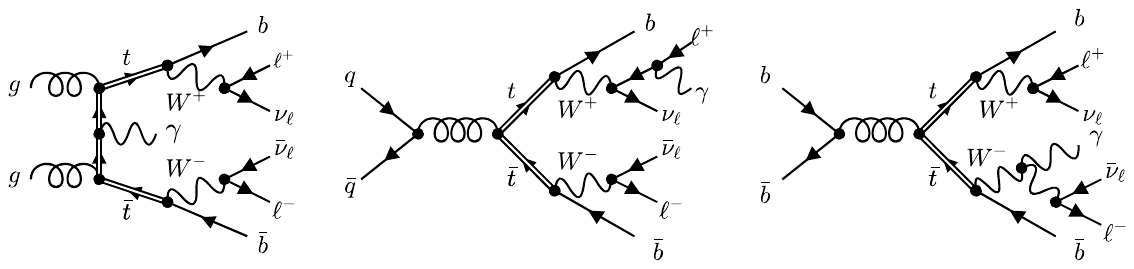}
 \end{center}
 \caption{\label{fig:fd-LO1} \it Example Feynman diagrams for $pp\to t\bar{t}\gamma$ contributing to \LOone. All Feynman diagrams in this paper were produced with the help of the \textsc{FeynGame} program \cite{Harlander:2020cyh}.}
\end{figure}

\begin{figure}[t!]
  \begin{center}
  \includegraphics[trim= 20 725 20 10, width=\textwidth]{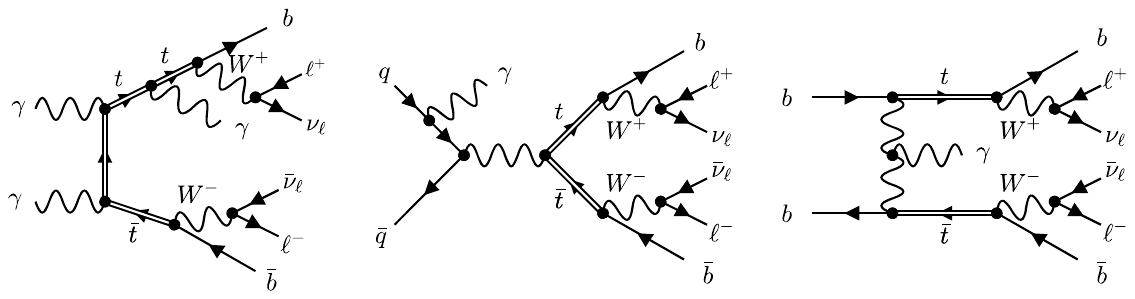}
\end{center}
  \caption{\label{fig:fd-LO3} \it Example Feynman diagrams for $pp\to t\bar{t}\gamma$ contributing to \LOthree.}
\end{figure}

At LO, both processes get contributions from three different orders of $\alpha_s$ and $\alpha$. The dominant contribution is at the order $\mathcal{O}(\alpha_s^2 \alpha^{4+n_{\gamma}})$, where $n_{\gamma}$ is the number of photons appearing in the Born-level process, and we call it \LOone, following the notation in Ref.~\cite{Frederix:2017wme,Frederix:2018nkq}. At this order, we encounter the typical QCD production of a top-quark pair, which leads to the following partonic subprocesses
\begin{equation}
\label{eq_lo1}
\begin{array}{c}
gg\to \ell^+\nu_{\ell}\, \ell^-\bar{\nu}_{\ell} \, 
b\bar{b}\,\gamma(\gamma) \,,\\[0.2cm]
\begin{array}{clllcll}
q\bar{q}/\bar{q}q&\to& \ell^+\nu_{\ell}\, 
\ell^{-}\bar{\nu}_{\ell} \, b\bar{b}\,\gamma(\gamma) \,,
&\eqspace&
b\bar{b}/\bar{b}b&\to& \ell^+\nu_{\ell}\, 
\ell^{-}\bar{\nu}_{\ell} \, b\bar{b}\,\gamma(\gamma) \,,
\end{array}
\end{array}
\end{equation}
with $q=u,d,c,s$. Example Feynman diagrams for the \LOone contribution are shown in Figure \ref{fig:fd-LO1}. We work in the five-flavour scheme and consistently include in the initial state all PDF suppressed channels from bottom quarks as well as photons in all subleading contributions. The \LOone contribution can be obtained from the square of matrix elements at the order $\mathcal{O}(g_s^2 g^{4+n_{\gamma}})$. The $q\bar{q}$ and $b\bar{b}$ subprocesses provide additional contributions to the amplitudes at the order $\mathcal{O}(g^{6+n_{\gamma}})$. The interference of both types of matrix elements gives rise to the first subleading LO contribution at $\mathcal{O}(\alpha_s^1 \alpha^{5+n_{\gamma}})$ which we refer to as \LOtwo.
\begin{figure}[t!]
  \begin{center}
  \includegraphics[trim= 20 710 20 10, width=\textwidth]{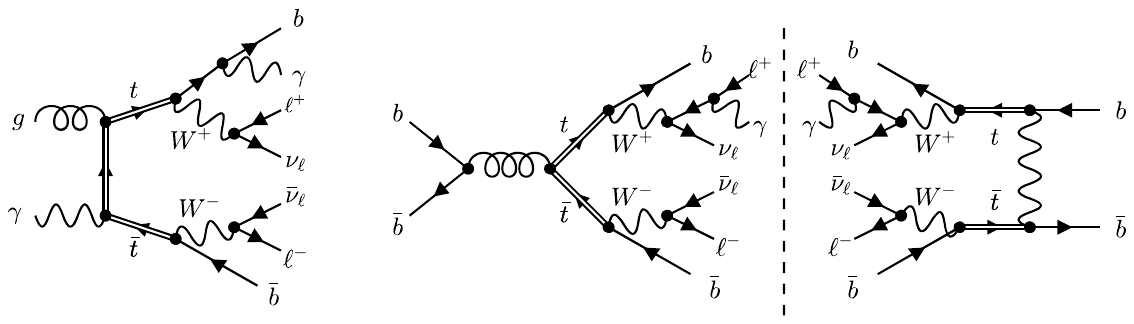}
\end{center}
  \caption{\label{fig:fd-LO2} \it Example Feynman diagrams for $pp\to t\bar{t}\gamma$ contributing to \LOtwo.}
\end{figure}
This interference is exactly zero for the $q\bar{q}$ initial state due to color algebra, but does not vanish for the $b\bar{b}$ subprocess due to additional $t$-channel Feynman diagrams with an intermediate $W$ boson, as shown on the right in Figure \ref{fig:fd-LO2}. Similar Feynman diagrams do not exist for $q\bar{q}$ because we keep the Cabibbo-Kobayashi-Maskawa (CKM) mixing matrix diagonal. In addition, at this order we encounter for the first time photon-induced channels with the $g\gamma$ initial state, as illustrated on the left in Figure \ref{fig:fd-LO2}. Thus, the partonic subprocesses of \LOtwo can be summarised as
\begin{equation}
\label{eq_lo2}
\begin{array}{c}
\begin{array}{clllcll}
g\gamma/\gamma g&\to& \ell^+\nu_{\ell}\, \ell^-\bar{\nu}_{\ell} \, 
b\bar{b}\,\gamma(\gamma) \,,
&\eqspace&
b\bar{b}/\bar{b}b&\to& \ell^+\nu_{\ell}\, 
\ell^{-}\bar{\nu}_{\ell} \, b\bar{b}\,\gamma(\gamma) \,.
\end{array}
\end{array}
\end{equation}
The third and last subleading LO contribution, \LOthree, is the purely EW induced production of a top-quark pair at the order $\mathcal{O}(\alpha^{6+n_{\gamma}})$. Example diagrams are depicted in Figure \ref{fig:fd-LO3}. Again, only the $q\bar{q}$ and $b\bar{b}$ channels are present, as well as the highly suppressed $\gamma\gamma$ channel. Therefore, the following reactions must be taken into account for \LOthree
\begin{equation}
\label{eq_lo3}
\begin{array}{c}
\gamma\gamma\to \ell^+\nu_{\ell}\, \ell^-\bar{\nu}_{\ell} \, 
b\bar{b}\,\gamma(\gamma) \,,\\[0.2cm]
\begin{array}{clllcll}
q\bar{q}/\bar{q}q&\to& \ell^+\nu_{\ell}\, 
\ell^{-}\bar{\nu}_{\ell} \, b\bar{b}\,\gamma(\gamma) \,,
&\eqspace&
b\bar{b}/\bar{b}b&\to& \ell^+\nu_{\ell}\, 
\ell^{-}\bar{\nu}_{\ell} \, b\bar{b}\,\gamma(\gamma) \,.
\end{array}
\end{array}
\end{equation}
With respect to \LOone, this contribution is not only suppressed by the power coupling, but also the gluon PDF does not enter this contribution at all. Finally, we denote as \LOfull the sum of all three LO contributions
\begin{equation}
{\rm LO} =  {\rm LO}_1 + {\rm LO}_2 + {\rm LO}_3\,.
\end{equation}

%
\subsection{NLO contributions}
\label{sec:desc-nlo}
%

%
\begin{figure}[t!]
  \begin{center}
  \includegraphics[trim= 20 580 20 20, width=0.6\textwidth]{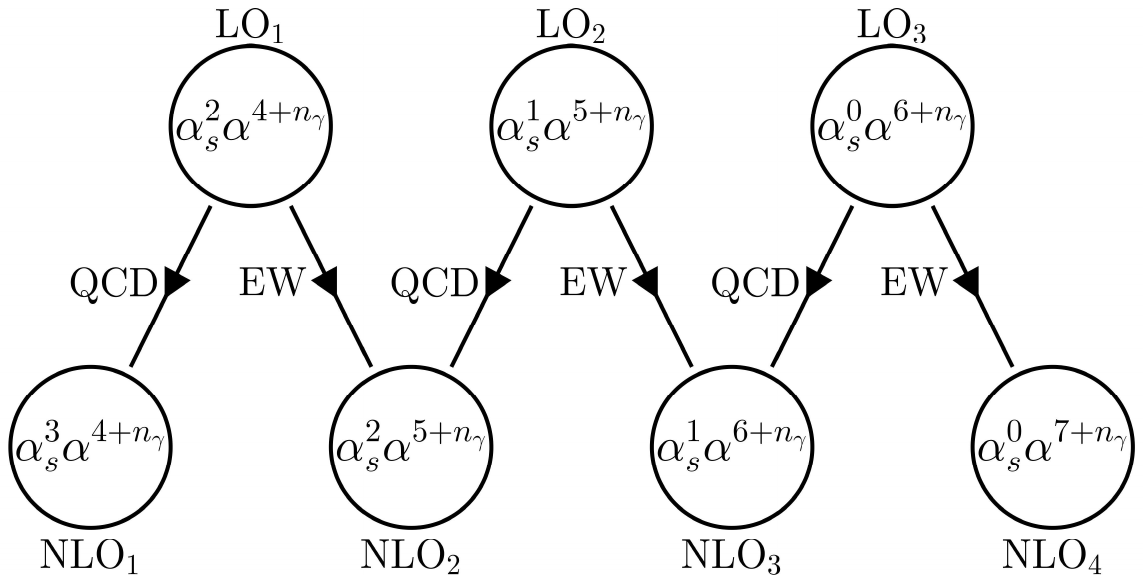}
\end{center}
  \caption{\label{fig:lo_nlo_cont} \it  LO and NLO contributions for $pp\to t\bar{t}\gamma(\gamma)$ with $n_\gamma=1(2)$.}
\end{figure}
\begin{figure}[t!]
  \begin{center}
  \includegraphics[trim= 20 720 20 10, width=\textwidth]{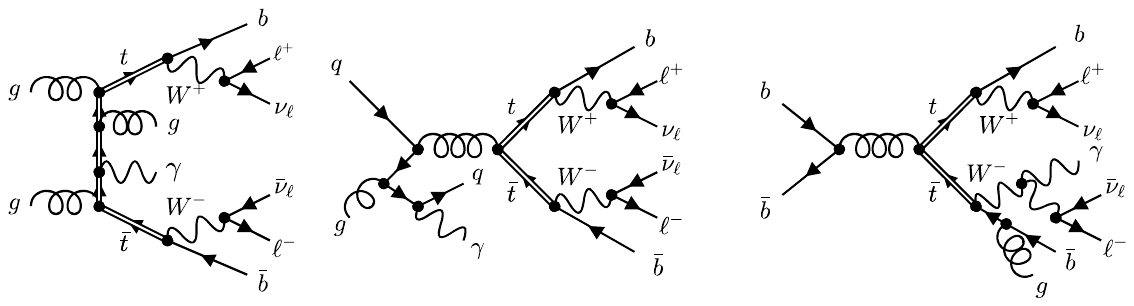}
\end{center}
  \caption{\label{fig:fd-NLO1} \it Example Feynman diagrams for $pp\to t\bar{t}\gamma$ contributing to \NLOone.}
\end{figure}

At NLO, we obtain corrections of QCD and/or EW nature to all LO contributions, as demonstrated in Figure \ref{fig:lo_nlo_cont}. The dominant higher-order corrections at NLO, denoted as \NLOone, arise from the QCD corrections to \LOone at the order $\mathcal{O}(\alpha_s^3 \alpha^{4+n_{\gamma}})$. We call the sum of the two \NLOqcd with 
\begin{equation}
{\rm NLO}_{\rm QCD}={\rm LO}_1+{\rm NLO}_1\,,
\end{equation}
and use it as a starting point to quantify the size of all subleading LO and NLO contributions. While the partonic subprocesses for the virtual corrections in \NLOone are identical to those of \LOone, for the real emission part additional reactions due to extra QCD radiation have to be taken into account. In particular, all partonic subprocesses can be obtained by adding an additional gluon and all possible crossings of partons, leading to
\begin{equation}
\label{eq_nlo1}
\begin{array}{c}
gg\to \ell^+\nu_{\ell}\, \ell^-\bar{\nu}_{\ell} 
\, b\bar{b}\,\gamma(\gamma)\, g \,,\\[0.2cm]
\begin{array}{clllcll}
q\bar{q}/\bar{q}q&\to& \ell^+\nu_{\ell}\, \ell^-\bar{\nu}_{\ell} 
\, b\bar{b}\,\gamma(\gamma)  \, g \,,
&\eqspace& 
b\bar{b}/\bar{b}b&\to& \ell^+\nu_{\ell}\, \ell^-\bar{\nu}_{\ell} 
\, b\bar{b}\,\gamma(\gamma)  \, g \,,    \\[0.2cm]
gq/qg &\to& \ell^+\nu_{\ell}\, \ell^-\bar{\nu}_{\ell} \, b\bar{b}\,\gamma(\gamma) \, q \,, 
&\eqspace& 
g\bar{q}/\bar{q}g&\to& \ell^+\nu_{\ell}\, \ell^-\bar{\nu}_{\ell} \, b\bar{b}\,\gamma(\gamma)  \, \bar{q}\,,\\[0.2cm]
gb/bg &\to& \ell^+\nu_{\ell}\, \ell^-\bar{\nu}_{\ell} \, b\bar{b}\,\gamma(\gamma) \, b \,,
&\eqspace& 
g\bar{b}/\bar{b}g&\to& \ell^+\nu_{\ell}\, \ell^-\bar{\nu}_{\ell} \, b\bar{b}\,\gamma(\gamma)  \, \bar{b}\,.
\end{array}
\end{array}
\end{equation}
Example Feynman diagrams for the real emission part of \NLOone are shown in Figure \ref{fig:fd-NLO1}. The real corrections also consistently include additional photon and gluon radiation in all stages, i.e. in the $t\bar{t}$ production and the decays of top quarks and $W$ gauge bosons. 

\begin{figure}[t!]
  \begin{center}
  \includegraphics[trim= 20 705 20 10, width=\textwidth]{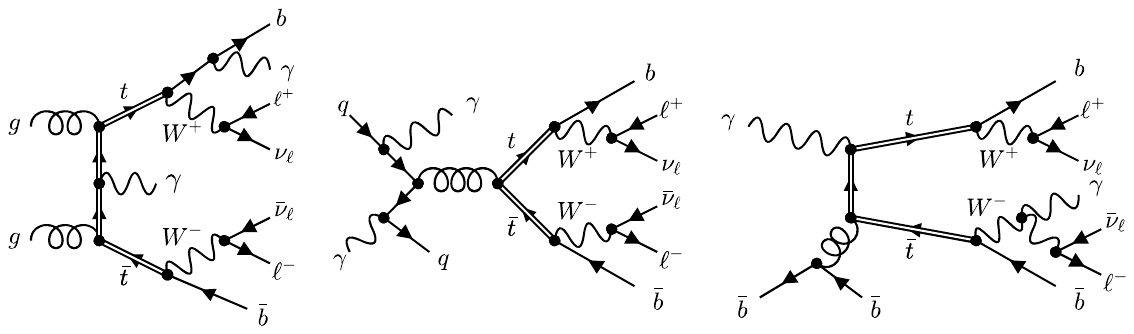}
\end{center}
  \caption{\label{fig:fd-NLO2} \it Example Feynman diagrams for $pp\to t\bar{t}\gamma$ contributing to \NLOtwo.}
\end{figure}

The next contribution, \NLOtwo, at the order $\mathcal{O}(\alpha_s^2 \alpha^{5+n_{\gamma}})$, cannot be completely separated into parts with only QCD or EW corrections. When calculating the corresponding virtual corrections we have on the one hand, the interference of tree-level diagrams at $\mathcal{O}(g^{6+n_{\gamma}})$ with one-loop diagrams at $\mathcal{O}(g_s^4g^{4+n_{\gamma}})$ and on the other hand, the interference between tree-level diagrams at $\mathcal{O}(g_s^2g^{4+n_{\gamma}})$ and one-loop diagrams at $\mathcal{O}(g_s^2g^{6+n_{\gamma}})$. The first contribution can be seen as a part of the NLO QCD corrections to \LOtwo. For the second contribution, no clear distinction is possible, since it can be seen as either the NLO QCD corrections to \LOtwo or the NLO EW corrections to \LOone. Finally, for the $g\gamma$ channel, we have the interference of tree-level diagrams at $\mathcal{O}(g_s^1g^{5+n_{\gamma}})$ with one-loop diagrams at $\mathcal{O}(g_s^3g^{5+n_{\gamma}})$. Compared to \NLOone, the real corrections in \NLOtwo include more partonic subprocesses, involving QCD and QED-like singularities. The first set of real corrections is obtained by additional QCD radiation to \LOtwo, which leads to
\begin{equation}
\label{eq_nlo2_1}
\begin{array}{c}
\begin{array}{clllcll}
q\bar{q}/\bar{q}q&\to& \ell^+\nu_{\ell}\, \ell^-\bar{\nu}_{\ell} 
\, b\bar{b}\,\gamma(\gamma)  \, g \,,
&\eqspace& 
b\bar{b}/\bar{b}b&\to& \ell^+\nu_{\ell}\, \ell^-\bar{\nu}_{\ell} 
\, b\bar{b}\,\gamma(\gamma)  \, g \,,    \\[0.2cm]
gq/qg &\to& \ell^+\nu_{\ell}\, \ell^-\bar{\nu}_{\ell} \, b\bar{b}\,\gamma(\gamma) \, q \,, 
&\eqspace& 
g\bar{q}/\bar{q}g&\to& \ell^+\nu_{\ell}\, \ell^-\bar{\nu}_{\ell} \, b\bar{b}\,\gamma(\gamma)  \, \bar{q}\,,\\[0.2cm]
gb/bg &\to& \ell^+\nu_{\ell}\, \ell^-\bar{\nu}_{\ell} \, b\bar{b}\,\gamma(\gamma) \, b \,,
&\eqspace& 
g\bar{b}/\bar{b}g&\to& \ell^+\nu_{\ell}\, \ell^-\bar{\nu}_{\ell} \, b\bar{b}\,\gamma(\gamma)  \, \bar{b}\,.
\end{array}
\end{array}
\end{equation}
In this case, we have the interference of tree-level diagrams at the order $\mathcal{O}(g_s^3g^{4+n_{\gamma}})$ and $\mathcal{O}(g_s^1g^{6+n_{\gamma}})$. Although the $q\bar{q}$ channel is exactly zero in \LOtwo, its NLO QCD corrections no longer vanish, and the corresponding virtual and real corrections must be taken properly into account. In particular, the interference of initial- and final-state gluon radiation gives rise to the non-vanishing contribution, as explained 
in Refs.~\cite{Denner:2016jyo,Denner:2023eti}. The latter contribution induces singularities that are  of soft nature only. Consequently, there are no collinear singularities in the $q\bar{q}$ channel and gluon radiation still vanishes in top-quark decays. It also follows that the $gq$ and $g\bar{q}$ channels contain no collinear singularities and turn out to be IR finite. Therefore, no dipole subtraction is required for these channels and the naive usage of it can even affect the efficiency of the phase-space integration. In the Nagy-Soper \cite{Bevilacqua:2013iha} subtraction scheme, which is used in our calculation, the corresponding collinear subtraction terms vanish by construction if the underlying Born-level matrix element is exactly zero, since they are directly proportional to it. This is, however, not a general feature of dipole subtraction schemes. For example, in the Catani-Seymour subtraction scheme \cite{Catani:1996vz,Catani:2002hc}, the corresponding subtraction term is a sum of the dipole terms $\mathcal{D}_k^{ai}$ involving color-correlated matrix elements. These dipole terms and the momentum mapping to the underlying Born-level process depend on the spectator parton $k$. Therefore, the sum of the dipole terms vanishes only in the exact collinear limit, where the dependence on the spectator parton $k$ vanishes and the sum becomes proportional to the Born-level matrix element. In principle, it is possible for such initial-state singularities to choose exactly one spectator parton and to substitute the color correlator $\mathbf{T}_k\cdot \mathbf{T}_{ai}\to -\mathbf{T}_{ai}^2$ in $\mathcal{D}_k^{ai}$. In this way the corresponding subtraction term is always proportional to the Born-level matrix element and would vanish as in the Nagy-Soper case. On the other hand, collinear singularities are present in the bottom-quark induced channels. In this case, gluon radiation in top-quark decays must be consistently included.  Furthermore, additional partonic subprocesses, which originate from \LOone with additional photon radiation, have to be included, leading to
\begin{equation}
\label{eq_nlo2_2}
\begin{array}{c}
gg\to \ell^+\nu_{\ell}\, \ell^-\bar{\nu}_{\ell} \, 
b\bar{b}\,\gamma\gamma(\gamma) \,,\\[0.2cm]
\begin{array}{clllcll}
q\bar{q}/\bar{q}q&\to& \ell^+\nu_{\ell}\, 
\ell^{-}\bar{\nu}_{\ell} \, b\bar{b}\,\gamma\gamma(\gamma) \,,
&\eqspace&
b\bar{b}/\bar{b}b&\to& \ell^+\nu_{\ell}\, 
\ell^{-}\bar{\nu}_{\ell} \, b\bar{b}\,\gamma\gamma(\gamma) \,.
\end{array}
\end{array}
\end{equation}
Similar to the calculation of NLO QCD corrections to  the $pp \to t\bar{t}j(j)$ process in the NWA \cite{Melnikov:2011qx,Bevilacqua:2022ozv}, photons can appear simultaneously in the production and decays, as illustrated in Figure \ref{fig:fd-NLO2}. Therefore, the relevant dipoles must be consistently included at each stage of the decay chain. Finally, we have new photon-induced partonic subprocesses which are constructed by gluon radiation from the $g\gamma$ channel in \LOtwo and by crossing from the processes listed in Eq. \eqref{eq_nlo2_2}, resulting in
\begin{equation}
\label{eq_nlo2_3}
\begin{array}{c}
g\gamma/\gamma g\to \ell^+\nu_{\ell}\, \ell^-\bar{\nu}_{\ell} \, b\bar{b}\,\gamma(\gamma)  \, g\,,\\[0.2cm]
\begin{array}{clllcll}
\gamma q/q\gamma &\to& \ell^+\nu_{\ell}\, \ell^-\bar{\nu}_{\ell} \, b\bar{b}\,\gamma(\gamma) \, q \,,
&\eqspace&
\gamma\bar{q}/\bar{q}\gamma&\to& \ell^+\nu_{\ell}\, \ell^-\bar{\nu}_{\ell} \, b\bar{b}\,\gamma(\gamma)  \, \bar{q}\,,\\[0.2cm]
\gamma b/b\gamma &\to& \ell^+\nu_{\ell}\, \ell^-\bar{\nu}_{\ell} \, b\bar{b}\,\gamma(\gamma) \, b \,,
&\eqspace&
\gamma\bar{b}/\bar{b}\gamma&\to& \ell^+\nu_{\ell}\, \ell^-\bar{\nu}_{\ell} \, b\bar{b}\,\gamma(\gamma)  \, \bar{b}\,.
\end{array}
\end{array}
\end{equation}
While the $g\gamma$ channel clearly contains only QCD singularities due to gluon radiation, for the remaining partonic processes we simultaneously encounter QCD and QED initial-state singularities arising from the $\gamma\to q\bar{q}$ and $q \to gq$ splittings, as shown in Figure \ref{fig:fd-NLO2}. This clearly illustrates that even real corrections cannot be divided into those of QCD or EW origin, as we have already seen in the calculation of virtual corrections.

\begin{figure}[t!]
  \begin{center}
  \includegraphics[trim= 20 700 20 10, width=\textwidth]{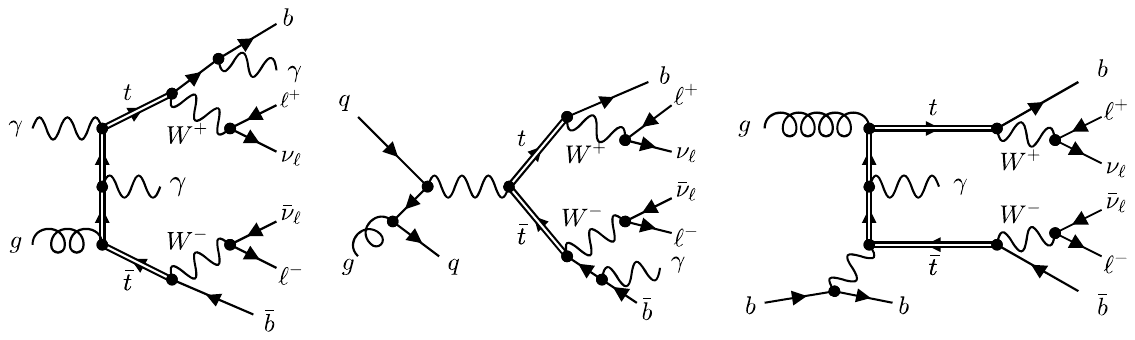}
\end{center}
  \caption{\label{fig:fd-NLO3} \it Example Feynman diagrams for $pp\to t\bar{t}\gamma$ contributing to \NLOthree.}
\end{figure}

The next subleading NLO contribution at the order $\mathcal{O}(\alpha_s^1 \alpha^{6+n_{\gamma}})$, \NLOthree, features a similarly wide range of virtual and real corrections with respect to \NLOtwo. This contribution is obtained from EW corrections to \LOtwo and QCD corrections to \LOthree. Thus, EW corrections to the $q\bar{q}$ channel in \LOtwo vanish again, since with respect to the Born level the color structure remains unchanged by these corrections. The virtual corrections consist of the interference of tree-level diagrams at $\mathcal{O}(g_s^2g^{4+n_{\gamma}})$ with the one-loop amplitude at $\mathcal{O}(g^{8+n_{\gamma}})$ and the tree-level matrix element at $\mathcal{O}(g^{6+n_{\gamma}})$ with one-loop diagrams at $\mathcal{O}(g_s^2g^{6+n_{\gamma}})$. The first contribution can be seen as part of the NLO EW corrections to \LOtwo, and thus it vanishes for $q\bar{q}$ as explained before, and only has to be included for the $b\bar{b}$ initial state. In the second term again no distinction between QCD and EW corrections is possible for the $q\bar{q}$ and $b\bar{b}$ channels, while for the $\gamma\gamma$ channel this contribution consists of pure QCD corrections to \LOthree. Finally, for $g\gamma$ we have the interference of tree-level diagrams at $\mathcal{O}(g_s^1g^{5+n_{\gamma}})$ with one-loop diagrams at $\mathcal{O}(g_s^1g^{7+n_{\gamma}})$. The real corrections can be summarised as follows. First, we have the NLO QCD corrections to \LOthree, which can be obtained by additional gluon radiation, bringing about the following contributions
\begin{equation}
\label{eq_nlo3_1}
\begin{array}{c}
\gamma\gamma\to \ell^+\nu_{\ell}\, 
\ell^{-}\bar{\nu}_{\ell} \, b\bar{b}\,\gamma(\gamma)g \,,\\[0.2cm]
\begin{array}{clllcll}
q\bar{q}/\bar{q}q&\to& \ell^+\nu_{\ell}\, 
\ell^{-}\bar{\nu}_{\ell} \, b\bar{b}\,\gamma(\gamma)g \,,
&\eqspace&
b\bar{b}/\bar{b}b&\to& \ell^+\nu_{\ell}\, \ell^-\bar{\nu}_{\ell} \, b\bar{b}\,\gamma(\gamma)  \, g\,.
\end{array}
\end{array}
\end{equation}
The crossing of initial- and final-state partons leads to the additional set of partonic subprocesses
\begin{equation}
\label{eq_nlo3_2}
\begin{array}{c}
\begin{array}{clllcll}
gq/qg &\to& \ell^+\nu_{\ell}\, \ell^-\bar{\nu}_{\ell} \, b\bar{b}\,\gamma(\gamma) \, q \,,
&\eqspace&
g\bar{q}/\bar{q}g&\to& \ell^+\nu_{\ell}\, \ell^-\bar{\nu}_{\ell} \, b\bar{b}\,\gamma(\gamma)  \, \bar{q}\,,\\[0.2cm]
gb/bg &\to& \ell^+\nu_{\ell}\, \ell^-\bar{\nu}_{\ell} \, b\bar{b}\,\gamma(\gamma) \, b \,,
&\eqspace&
g\bar{b}/\bar{b}g&\to& \ell^+\nu_{\ell}\, \ell^-\bar{\nu}_{\ell} \, b\bar{b}\,\gamma(\gamma)  \, \bar{b}\,,
\end{array}
\end{array}
\end{equation}
which again contain simultaneously QCD and QED initial-state singularities, as illustrated in Figure \ref{fig:fd-NLO3}. Next, we obtain additional contributions from photon radiation off the partonic subprocesses in \LOtwo, as exemplified in Figure \ref{fig:fd-NLO3}. Together with all possible crossings of partons and photons we obtain
\begin{equation}
\label{eq_nlo3_3}
\begin{array}{c}
\begin{array}{clllcll}
g\gamma/\gamma g&\to& \ell^+\nu_{\ell}\, 
\ell^{-}\bar{\nu}_{\ell} \, b\bar{b}\,\gamma\gamma(\gamma) \,
&\eqspace&
b\bar{b}/\bar{b}b&\to& \ell^+\nu_{\ell}\, 
\ell^{-}\bar{\nu}_{\ell} \, b\bar{b}\,\gamma\gamma(\gamma) \,,\\[0.2cm]
\gamma b/b\gamma &\to& \ell^+\nu_{\ell}\, \ell^-\bar{\nu}_{\ell} \, b\bar{b}\,\gamma(\gamma) \, b \,
&\eqspace&
\gamma\bar{b}/\bar{b}\gamma&\to& \ell^+\nu_{\ell}\, \ell^-\bar{\nu}_{\ell} \, b\bar{b}\,\gamma(\gamma)  \, \bar{b}\,.
\end{array}
\end{array}
\end{equation}
In addition, $q\bar{q}$ and $\gamma q/\gamma \bar{q}$ subprocesses vanish due to color, as additional photon radiation does not change the color structure already present in \LOtwo.

Since in \NLOthree we find for the first time EW 1-loop corrections in photon-initiated subprocesses, in particular in the $g\gamma$ channel, we briefly mention here the choice of the renormalisation scheme for the electromagnetic coupling $\alpha$, which will be explained in detail later in the article. As  we are interested in the experimental signature with at least one (two) isolated photon(s), we do not consider the $\gamma\to f\bar{f}$ splitting for final state photons. Therefore, the electromagnetic coupling constant connected with final state photons has to be renormalised in the on-shell scheme to obtain IR finite results, see e.g. \ Refs.~\cite{Denner:2019vbn,Pagani:2021iwa}. However, the electromagnetic coupling associated with initial state photons must be renormalised in a $\overline{\rm MS}$-like scheme, because $\gamma\to f\bar{f}$ splittings are included in the evolution of the proton PDF. 
Thus, we use the $G_\mu$ scheme in this case and the remaining IR $\epsilon$ poles from the sum of real and virtual corrections are absorbed into a redefinition of PDFs, as for all other QCD and QED collinear initial state singularities. This leads to a collinear factorisation counterterm of the form
\begin{equation}
\frac{\alpha}{2\pi}\frac{1}{\Gamma (1-\epsilon)} \frac{1}{\epsilon}\left(\frac{4\pi \mu_R^2}{\mu_F^2} \right)^\epsilon P_{\gamma\gamma}(x),
\end{equation}
which has to be convoluted with the corresponding Born-level cross section. The one-loop Altarelli-Parisi QED kernel $P_{\gamma\gamma}(x)$, see e.g. \ Ref.~\cite{Frederix:2018nkq} for more details, is given by
\begin{equation}
P_{\gamma\gamma}(x)=-\frac{2}{3}\,\delta (1-x)\,\sum_f N_{c,f}\,Q_f^2,
\end{equation}
where the summation runs over all massless, charged fermions, $N_{c,f}$ is a color factor, which is $3$ for quarks and $1$ for leptons, and $Q_f$ is the charge of the fermion $f$. Unlike all other QCD and QED splitting functions, $P_{\gamma\gamma}(x)$ contains only a virtual contribution given by the delta distribution.

\begin{figure}[t!]
  \begin{center}
  \includegraphics[trim= 20 725 20 10, width=\textwidth]{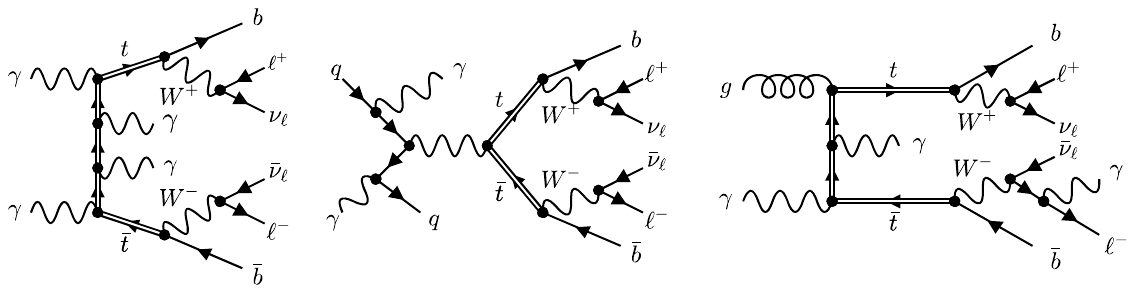}
\end{center}
  \caption{\label{fig:fd-NLO4} \it Example Feynman diagrams for $pp\to t\bar{t}\gamma$ contributing to \NLOfour.}
\end{figure}

Finally, we have the NLO EW corrections to \LOthree, called \NLOfour, and all new partonic subprocesses can simply be obtained from additional photon radiation and crossing, leading to
\begin{equation}
\label{eq_nlo4}
\begin{array}{c}
\gamma\gamma\to \ell^+\nu_{\ell}\, \ell^-\bar{\nu}_{\ell} 
\, b\bar{b}\,\gamma\gamma(\gamma)\,,\\[0.2cm]
\begin{array}{clllcll}
q\bar{q}/\bar{q}q&\to& \ell^+\nu_{\ell}\, \ell^-\bar{\nu}_{\ell} 
\, b\bar{b}\,\gamma\gamma(\gamma)\,,
&\eqspace& 
b\bar{b}/\bar{b}b&\to& \ell^+\nu_{\ell}\, \ell^-\bar{\nu}_{\ell} 
\, b\bar{b}\,\gamma\gamma(\gamma)\,,    \\[0.2cm]
\gamma q/q\gamma &\to& \ell^+\nu_{\ell}\, \ell^-\bar{\nu}_{\ell} \, b\bar{b}\,\gamma(\gamma) \, q \,, 
&\eqspace& 
\gamma\bar{q}/\bar{q}\gamma&\to& \ell^+\nu_{\ell}\, \ell^-\bar{\nu}_{\ell} \, b\bar{b}\,\gamma(\gamma)  \, \bar{q}\,,\\[0.2cm]
\gamma b/b\gamma &\to& \ell^+\nu_{\ell}\, \ell^-\bar{\nu}_{\ell} \, b\bar{b}\,\gamma(\gamma) \, b \,,
&\eqspace& 
\gamma\bar{b}/\bar{b}\gamma&\to& \ell^+\nu_{\ell}\, \ell^-\bar{\nu}_{\ell} \, b\bar{b}\,\gamma(\gamma)  \, \bar{b}\,,
\end{array}
\end{array}
\end{equation}
where example Feynman diagrams are shown in Figure \ref{fig:fd-NLO4}.

We denote the complete result at NLO, including all subleading contributions, as \NLOfull. The latter result is given by the following sum
\begin{equation} \label{eq:nlo_tot}
{\rm NLO}={\rm LO}_1+{\rm LO}_2+{\rm LO}_3+{\rm NLO}_1+{\rm NLO}_2+{\rm NLO}_3+{\rm NLO}_4.
\end{equation}
We refrain from dividing our complete NLO calculation into different resonant contributions based on the origin of photon emission, as already done at NLO QCD for the $pp\to t\bar{t}\gamma$ process in Refs.~\cite{Melnikov:2011ta,Bevilacqua:2019quz} and for $pp\to t\bar{t}\gamma \gamma$ in Ref.~\cite{Stremmer:2023kcd}. It is because additional photon radiation introduces further complications due to mixing of Born-level processes with different photon radiation patterns at LO. This issue has already been examined in detail in the case of the calculation of NLO QCD corrections to $pp\to t\bar{t}jj$ \cite{Bevilacqua:2022ozv}. Instead, we perform an alternative calculation, labelled as \NLOprd, which we define according to
\begin{equation}
{\rm NLO}_{\rm prd}={\rm LO}_1+{\rm LO}_2+{\rm LO}_3+{\rm NLO}_1+{\rm NLO}_{2,{\rm prd}}+{\rm NLO}_{3,{\rm prd}}+{\rm NLO}_{4,{\rm prd}},
\end{equation}
where the subscript \textit{prd} indicates that photon bremsstrahlung and subleading NLO corrections are only included in the production stage of the $pp\to t\bar{t}\gamma(\gamma)$ process. In detail, in \NLOprd, all LO (\LOone, \LOtwo, \LOthree) contributions as well as \NLOone are fully included in both the $t\bar{t}\gamma(\gamma)$ production and decays of the top-quark pair. On the other hand, the subleading NLO corrections (\NLOtwo, \NLOthree, \NLOfour) are only included in the production stage of the $t\bar{t}\gamma (\gamma)$ process when the photon (the two photons) are also present there. In this way, we investigate the origin of the main subleading corrections. This approximation is motivated by the fact that, on the one hand, it leads to an enormous simplification of the calculations, especially for the real emission part, and ultimately in the matching to parton shower programs. On the other hand, the largest subleading NLO contributions are expected to originate from EW Sudakov logarithms in the tails of dimensionful observables. However, it is known that high $p_T$ phase-space regions are dominated by the case where all photons are produced in the $t\bar{t}$ production \cite{Melnikov:2011ta,Bevilacqua:2019quz,Stremmer:2023kcd}, which should directly reduce the relevance of including subleading NLO corrections in other radiation patterns. With the calculation of ${\rm NLO}_{\rm prd}$ we would like to confirm that this is the case for the $pp\to t\bar{t}\gamma (\gamma)$ process for various differential cross-section distributions. For the sake of completeness, we  will also show the ${\rm NLO}_{\rm prd}$ result for the $t\bar{t}\gamma(\gamma)$ integrated fiducial cross section. 

%
\section{Computational framework}
\label{sec:description}
%

As we have already alluded to in Section \ref{sec:definition} the calculation of real corrections is performed with the Nagy-Soper subtraction scheme \cite{Bevilacqua:2013iha}, which has recently been extended to NLO QCD calculations in the NWA involving internal on-shell resonances \cite{Bevilacqua:2022ozv}. To check the validity of the results, we also use the Catani-Seymour subtraction scheme \cite{Catani:1996vz,Catani:2002hc}. Furthermore, for additional cross-checks in both cases we utilize the phases-space restriction on the subtraction terms \cite{Nagy:1998bb,Nagy:2003tz,Bevilacqua:2009zn,Czakon:2015cla}. For the purposes of this study, both subtraction schemes, that are implemented in the \helacdipoles MC program \cite{Czakon:2009ss}, have been extended to deal with soft and collinear singularities of QED origin that plague the real radiation contributions of NLO EW calculations. In the following we describe both extensions in detail and discuss how the calculation of virtual corrections is organised for the purposes of our study.

\subsection{Real corrections}

At NLO QCD, we have to deal with contributions of different origin, on the one hand with virtual corrections due to the interference of one-loop and tree-level amplitudes, and on the other hand with real corrections arising from the emission of additional QCD partons. However, both contributions themselves are divergent, and only the sum of them is finite for IR safe observables. A standard approach nowadays is the use of a subtraction scheme, where an auxiliary cross section is introduced, which is designed to locally mimic the singular limits of the real-correction cross section. Moreover, this auxiliary cross section has to be simple enough so that 
the integration over the one-parton subspaces, which cause the soft and collinear divergences, is possible either analytically or numerically, leading to the so-called integrated subtration terms and/or dipoles. These can then be combined with the virtual corrections to cancel all IR divergences. In general, the subtraction term in QCD can be written as
\begin{equation} \label{eq:dipole}
\mathcal{A}^{\textrm{D}}=\sum_{i,j,k=1}^{n+1} \mathcal{A}^{\textrm{B}}(\{\tilde{p}\}_n^{ijk}) \otimes\mathcal{D}^{(ijk)}(\{\tilde{p}\}_n^{(ijk)},\{p\}_{n+1}) \, \left(\mathbf{T}_{ij} \cdot \mathbf{T}_k\right),
\end{equation}
where $\mathcal{A}_{\textrm{B}}$ is the tree-level matrix element of the underlying Born-level process which results from the recombination of the two partons $i$ and $j$. The terms $\mathcal{D}^{(ijk)}$ are the so-called dipoles and $\{\tilde{p}\}_n^{(ijk)}$ is the new set of momenta for the underlying Born-level process, which is obtained by a momentum mapping from the initial momenta $\{p\}_{n+1}$. Spin correlations between the squared Born-level matrix element and the dipoles are indicated by the symbol $\otimes$. Finally, $\mathbf{T}_k$ are color operators according to the definitions given in Ref.~\cite{Catani:1996vz}. The indices $i,j,k$ indicate that in general the dipoles as well as the momentum mapping depend on the two splitting partons $i,j$ and potentially also on the spectator parton $k$. The QED subtraction scheme can be obtained from the QCD one by simple substitutions. In particular, the color operators have to be replaced by charge operators according to
\begin{equation}
\mathbf{T}_{ij} \cdot \mathbf{T}_k \rightarrow \mathbf{Q}_{ij} \mathbf{Q}_k,
\end{equation} 
where $\mathbf{Q}_{k}$ is the charge of the particle with a relative minus sign between final- and initial-state particles, and possible further simple replacements of color factors. However, this cannot be applied to the case where the splitting pair is recombined into a photon ($Q_{ij}=0$). Instead, in this case we perform the following substitution
\begin{equation}
\mathbf{T}_{ij} \cdot \mathbf{T}_k \rightarrow -w_k\mathbf{Q}_{i}^2\;\;\;\;\;\;\;\;\textrm{with}\;\;\;\;\;\;\;\; \sum_k w_k=1.
\end{equation} 
The particular choice of these weights $w_k$ is completely arbitrary and, in our case we simply set it to $\delta_{k,k_0}$, which implies that we choose exactly one spectator particle. In general, this spectator particle must be the same in the actual subtraction and in the calculation of the integrated dipoles, since the momentum mapping and dipole terms depend on $k$. We note that in the Nagy-Soper subtraction scheme, both the momentum mapping and the collinear dipole terms are independent of the spectator particle, so that the choice of $k_0$ plays no role in the calculation. Finally, the structure of \helacdipoles has been modified to simultaneously keep track of different orders in $\alpha_s$ and $\alpha$ and to allow the reweighting to different renormalisation and factorisation scale settings on the fly. To verify the new modifications, we have reproduced the results of 
Ref.~\cite{Denner:2016jyo} at the integrated and differential cross section level. In detail, we have calculated afresh NLO EW corrections to the  $pp \to t\bar{t}+X$  process in the di-lepton top-quark decay channel for the LHC Run II center-of-mass energy of $\sqrt{s}=13$ TeV.   In this calculation all finite-width effects, resonant and non-resonant contributions of the top quark and $W^\pm/Z$ gauge bosons as well as interference effects have been included.

Next, we discuss the changes in the Nagy-Soper subtraction scheme for calculations involving resonant particles. These modifications are to a large extent independent of QCD or QED-like singularities. To this end, we  first briefly recall the notation and main aspects of the Nagy-Soper subtraction scheme which were presented in detail in  Ref.~\cite{Bevilacqua:2013iha}. We focus on the case of final-state splittings  since initial-state splittings remain unaffected.  We  first introduce the following definitions
\begin{gather} \label{ns:def}
Q=p_1+p_2=\sum_{l=3}^{n+1} p_l,\\
P_{ij}=p_i+p_j\qquad\qquad\textrm{and}\qquad\qquad
K=Q-P_{ij},
\end{gather}
where $Q$ is the total momentum of the process, $P_{ij}$ is the momentum sum of the splitting partons and $K$ is the so-called collective spectator. The momentum mapping from $\{p\}_{n+1}$ to $\{\tilde{p}\}_n^{(ij)}$ is then defined by the condition that the momentum of the splitting parton $\tilde{p}_i$ should lie in the $Q$-$P_{ij}$ plane according to
\begin{equation} \label{eq:pij}
P_{ij}=\beta \tilde{p}_i+\gamma Q,
\end{equation}
where  $\beta$ and $\gamma$ are uniquely fixed by the following two conditions
\begin{align} \label{eq:ns_map}
\tilde{Q}&=Q \,, \\
\widetilde{K}^2&=K^2\,.
\end{align}
From Eq.~\eqref{eq:ns_map} it is clear that the momentum mapping from $K$ to $\widetilde{K}$ is given by a Lorentz transformation ($k_i^{\mu}=\tensor{\Lambda(K,\widetilde{K})}{^\mu_\nu}\,\widetilde{k}_i^{\nu}$), which is defined as \cite{Nagy:2007ty,Bevilacqua:2013iha}
\begin{equation}
\tensor{\Lambda(K,\widetilde{K})}{^\mu_\nu}=\tensor{g}{^{\mu}_{\nu}}-\frac{2(K+\widetilde{K})^\mu(K+\widetilde{K})_\nu}{(K+\widetilde{K})^2}+\frac{2K^\mu\widetilde{K}_\nu}{K^2}.
\end{equation}
Next, the dipole terms $\mathcal{D}^{(ijk)}$ in Eq.~\eqref{eq:dipole} are decomposed into diagonal terms $W^{(ii,j)}$ containing soft and collinear singularities and interference terms $W^{(ik,j)}$ including only soft singularities according to 
\begin{equation}
\mathcal{D}^{(ijk)}_{\tilde{s}_1\tilde{s}_2}=W^{(ii,j)}_{\tilde{s}_1\tilde{s}_2}\delta_{ik}+W^{(ik,j)}_{\tilde{s}_1\tilde{s}_2}(1-\delta_{ik})\,,
\end{equation}
where $\tilde{s}_1$ and $\tilde{s}_2$ are the helicity values of the splitting parton. The diagonal and interference terms are further split into a sum of so-called splitting functions $v^{(ij)}$ as specified by
\begin{align} \label{eq:ns_splitt}
W^{(ii,j)}_{\tilde{s}_1\tilde{s}_2}&=\sum_{s_i,s_j}v^{(ij)}_{\tilde{s}_1s_is_j}\left(v^{(ij)}_{\tilde{s}_2s_is_j}\right)^*, \\
W^{(ik,j)}_{\tilde{s}_1\tilde{s}_2}&=\sum_{s_i,s_j}v^{(ij),eik}_{\tilde{s}_1s_is_j}\left(v^{(kj),eik}_{\tilde{s}_2s_is_j}\right)^* \,,
\end{align}
where $v^{(ij),eik}$ is the eikonal approximation of these splitting functions. In QCD the $v^{(ij),eik}$ function reduces to a much simpler expression given by 
\begin{align}
v^{(ij),eik}_{\tilde{s}_is_is_j}=\sqrt{4\pi\alpha_s}\,\delta_{\tilde{s}_is_i}\frac{\epsilon(p_j,s_j)^*\cdot p_i}{p_i\cdot p_j}.
\end{align}
The splitting functions are obtained by factorising the splitting $\tilde{p}_i\to p_i + p_j$ from the divergent matrix element in the following way $\mathcal{M}_{n+1}^{\rm div}=\mathcal{M}_n \otimes v^{(ij)}$. The exact form for the different QCD splittings for final- and initial-state splittings can be found in Ref.~\cite{Nagy:2007ty}.

The original Nagy-Soper subtraction scheme can directly be used in the case of unresolved partons in the production process. However, if an unstable particle is part of the emitter pair, then we use the spin-averaged versions of the dipole terms $\mathcal{D}^{(ijk)}$ in Eq.~\eqref{eq:dipole}. In addition, we use the eikonal approximation for the splitting functions in the diagonal term $W^{(ii,j)}$ given in Eq.~\eqref{eq:ns_splitt} when a $W$ gauge boson is part of the emitter pair. These modifications are also applied to unstable particles in nested decay chains, such as $W$ gauge bosons in radiative top-quark decays. Further modifications are required for the dipole subtraction in decay processes. The first change concerns the momentum mapping which is required to preserve the total momentum of the decay process, i.e. the momentum of the decaying mother particle. This is achieved by replacing the total momentum $Q$ in the original momentum mapping with the momentum of the mother particle, which is called $Q_{\rm dec}$ in analogy. By definition $Q_{\rm dec}$ is then preserved under the momentum mapping according to the condition defined in Eq.~\eqref{eq:ns_map}. The dipole terms $\mathcal{D}^{(ijk)}$ also have an explicit and an additional implicit dependence on $Q$ by the axial gauge with $Q$ as reference vector used for the polarisation vectors of gluons and photons. Thus, to reuse all the results of the original subtraction scheme, we have to perform the replacement $Q\to Q_{\rm dec}$ also in all the dipole terms $\mathcal{D}^{(ijk)}$ and apply a gauge transformation to the external polarisation vectors entering the splitting functions $v^{(ij)}$. The mother particle is excluded from the list of possible emitters in the dipole subtraction, but it is still included as a possible spectator particle. To properly account for all soft singularities associated with the mother particle, we replace $W^{(ik,j)}$, defined in Eq.~\eqref{eq:ns_splitt}, in the case where the mother particle is the spectator $k$, by $W^{(ik,j)}_{\rm dec}$, which we define as
\begin{align} \label{eq:ns_wdec}
W_{\rm dec}^{(ik,j)}=W^{(ik,j)}+W^{(ki,j)}-W^{(kk,j),eik}.
\end{align}
For the calculation of the corresponding integrated dipoles we closely follow the semi-numerical approach 
of Ref.~\cite{Bevilacqua:2013iha}. In this case, the spin-averaged diagonal and interference terms are cast in the following form
\begin{align} 
\overline{W}^{(ii,j)} - \overline{W}_{ \rm dec}^{(ik,j)} = \underbrace{\left( \overline{W}^{(ii,j)} -\overline{W}^{(ii,j),eik}\right)}_{\mathcal{D}_{ii}} + \underbrace{\left(\overline{W}^{(ii,j),eik} -  \overline{W}_{\rm dec}^{(ik,j)}\right)}_{\mathcal{I}_{ik,{\rm dec}}},
\end{align}
where $\mathcal{D}_{ii}$ can only have collinear singularities by construction and are already known from the original formulation. The second term can have soft and collinear singularities and can be further simplified to 
\begin{align} \label{eq:ns_Iik_dec}
\mathcal{I}_{ik,{\rm dec}}=4\pi\alpha_s \frac{-((p_j\cdot p_k)p_i-(p_i\cdot p_j)p_k)^2}{(p_i\cdot p_j)^2(p_j\cdot p_k)^2}.
\end{align}
We have calculated and implemented this new dipole in the case of massless and massive splitting partons $\tilde{p}_i$. The latter case is required for the calculation of EW corrections to top-quark decays, i.e. for $W^{\pm}\to W^{\pm}\gamma$ splittings. In addition, it can also be used for e.g. NLO QCD corrections with massive bottom quarks inside top-quark decays. In summary, our implementation of the Nagy-Soper subtraction scheme allows us to calculate QCD and EW radiative corrections for any number of coloured, charged, massive and massless particles in nested decay chains.

We have carried out several tests to confirm the correctness of the modifications we have made. First of all, we have already used this new extension for the calculation of the NLO QCD corrections to the $pp\to t\bar{t}jj$ process in the di-lepton top-quark decay channel in the NWA  \cite{Bevilacqua:2022ozv}. These results, were cross-checked with an alternative calculation using the Catani-Seymour subtraction scheme and its extension to the NWA case \cite{Campbell:2004ch,Melnikov:2011qx,Bevilacqua:2019quz}. In addition, we have recalculated the NLO QCD corrections of the top-quark width with massive bottom quarks \cite{Jezabek:1988iv,Czarnecki:1990kv} as well as  the corresponding real corrections, which are presented in the appendix of Ref.~\cite{Campbell:2012uf}. Finally, we have reproduced the calculation of the NLO electroweak corrections to the top-quark decay width while treating the internal $W$ boson in the NWA \cite{Basso:2015gca}.

In our current calculation of the $pp\to t\bar{t}\gamma(\gamma)$ process in the NWA,  we have computed the real correction part of the \NLOfull calulations with the Nagy-Soper subtraction scheme. We  have utilised two different values of the phase-space restriction on the subtraction terms and found excellent agreement for the obtained results within the corresponding MC errors. Finally, we have reproduced the results of ${\rm NLO}_{i,{\rm prd}}$ with $i=1,2,3,4$ using the Catani-Seymour subtraction scheme. A full comparison with the Catani-Seymour subtraction is not yet possible as our implementation is currently limited to massless emitters in decay processes.

\subsection{Virtual corrections}

The  calculation of the virtual corrections is organized using reweighting techniques \cite{Binoth:2008kt,Bevilacqua:2009zn}. In practice, partially unweighted events \cite{Bevilacqua:2016jfk} that are generated at the Born level, are then used to calculate the one-loop corrections. The phase-space integration is performed using the programs \textsc{Parni} \cite{vanHameren:2007pt} and \textsc{Kaleu} \cite{vanHameren:2010gg}. The  combined tree-level and one-loop results are then stored in modified Les Houches Event Files (LHEFs) \cite{Alwall:2006yp,Bern:2013zja}, which have been extended to save, for each event, the necessary information for all contributions at different orders of $\alpha_s$ and $\alpha$. This allows us, among other things, to reweight our results to different renormalisation and factorisation scale settings. Tree-level and one-loop matrix elements are obtained with the help of the \Recola  matrix element generator \cite{Actis:2016mpe,Actis:2012qn}. We have modified the calculation of the matrix elements in \Recola to incorporate the random polarisation method \cite{Draggiotis:1998gr,Draggiotis:2002hm,Bevilacqua:2013iha} which leads to significant performance improvements in the phase-space integration. Following the notation of Ref.~\cite{Actis:2016mpe}, every one-loop  amplitude can be written as a linear combination of tensor integrals
\begin{equation}
   \mathcal{A}_{1-{\rm loop}} = \sum_t c_{\mu_1 \dots \mu_{r_t}}^{(t)} \, T^{\mu_1 \dots \mu_{r_t}}_{(t)}+\mathcal{A}_{\rm CT}\,,
\end{equation}
where $\mathcal{A}_{\rm CT}$ include the counterterms, $c_{\mu_1 \dots \mu_{r_t}}^{(t)}$ are the tensor coefficients that do not depend on the loop momentum $q$, $t$ classifies the different tensor integrals needed for the process and $r_t$ labels the rank of the tensor integral. Furthermore, the tensor integrals are given by 
\begin{equation}
T^{\mu_1 \dots \mu_{r_t}}_{(t)} = \frac{(2\pi\mu)^{4-D}}{i\pi^2} \int d^Dq \frac{q^{\mu_1}\dots q^{\mu_{r_t}}}{D_{0}^{(t)} \dots D_{k_t}^{(t)}} \,,  \quad \quad \quad \quad D_{i}^{(t)} = (q+p_i^{(t)})^2-(m_{i}^{(t)})^2\,,
\end{equation}
where $k_t$ is the number of propagators. The tensor coefficients are calculated by \Recola  similarly to the tree-level amplitudes by recursive construction of loop off-shell currents \cite{vanHameren:2009vq,Actis:2012qn}. The relevant tensor integrals are computed with the \textsc{Collier} library \cite{Denner:2016kdg} which implements the reduction techniques of \cite{Denner:2005nn,Denner:2002ii} and the scalar integrals of \cite{Denner:2010tr}. Alternatively, we have implemented the reduction to scalar integrals with \textsc{CutTools} \cite{Ossola:2007ax}. The resulting scalar integrals are computed with \textsc{OneLOop} \cite{vanHameren:2010cp}. The \textsc{CutTools} program implements the OPP reduction method \cite{Ossola:2006us}, which is based on double, triple and quadruple cuts. A similar approach has already been employed in the \textsc{OpenLoops} program \cite{Buccioni:2017yxi}. In our case, the OPP reduction and the evaluation of the scalar integrals are performed with quadruple precision, while the tensor coefficients are calculated with double precision.  Not only do we use this second reduction scheme for additional cross-checks of our framework, but also for phase-space points containing tensor integrals that have been flagged by \textsc{Collier} as possibly unstable. 

As explained in Ref.~\cite{Actis:2016mpe} the \textsc{Recola} program offers the possibility to select the resonant parts of the amplitude, and can therefore be employed to calculate matrix elements in the pole approximation. In this approximation, only the resonant parts of the amplitude are kept, thus, the complex squared mass  of the resonant particle $(\mu^2= m^2 -i\Gamma m)$ is replaced by its real part $(\Re(\mu^2) = m^2)$ everywhere in the amplitude except for the denominators of resonant propagators where the width $\Gamma$ of the unstable particle is still present. The NWA matrix element is then obtained by performing the usual limit $\Gamma/m\to 0$ in the Breit-Wigner propagators. Finally, we correct the symmetry factor calculated in \Recola for processes involving identical particles in the final state but originating from  different subprocesses in the decay chain, as explained in Ref.~\cite{Denner:2020orv}.

As an alternative, we have implemented a second fully automatic method for the construction of one-loop matrix elements in the NWA, which is based on the calculation of the on-shell amplitudes for the individual contributions needed in e.g. the $pp \to t\bar{t} (\gamma \gamma)$ process in the di-lepton top-quark decay channel. To this end, we compute separately the subamplitudes for  $pp \to t\bar{t} (\gamma\gamma)$, $t\to W^+b (\gamma\gamma)$, $\bar{t}\to W^- \bar{b}(\gamma\gamma)$, $W^+\to \ell^+\nu_\ell (\gamma\gamma)$ and $W^-\to \ell^- \bar{\nu}_\ell (\gamma\gamma)$  with \Recola and  then combine them appropriately in color and helicity space to obtain the full one-loop amplitude. This alternative implementation is particularly useful for cross-checking because in this case we can switch on and off one-loop corrections for specific contributions. We note here, that we have already used both approaches in Ref.~\cite{Bevilacqua:2022ozv} where  NLO QCD calculations  for the $pp\to t\bar{t}jj$ process in the di-lepton top-quark decay channel have been carried out. 

We have performed  various cross-checks between \Recola and \textsc{Recola2} \cite{Denner:2017wsf} for all partonic subprocesses.  Specifically, for each subprocess and for several phase-space points we checked the finite remainders as well as the coefficients of the poles in $\epsilon$ of the virtual amplitudes. For testing purposes we utilised the Background-Field-Method (BFM) as implemented in \textsc{Recola2} \cite{Denner:1994xt,Denner:2017vms,Denner:2019vbn}.  The BFM,  which is an alternative formulation for the quantisation of gauge fields that can be seen as a different gauge, can be used as a complementary method  in the calculation of one-loop matrix elements besides the usual formulation.

The proper description of unstable particles like $W^\pm$ gauge bosons and top quarks in perturbation theory requires the introduction of the complex-mass scheme (CMS), which was first proposed for lowest-order calculations in 
Ref.~\cite{Denner:1999gp} and then generalised  to higher orders in Refs.~\cite{Denner:2005fg,Denner:2006ic}. In this scheme, complex masses are introduced at the level of the Lagrangian by splitting the bare masses into complex renormalised masses and complex counterterms. This corresponds to a rearrangement of the perturbative expansion without changing the underlying theory and preserving gauge invariance. Unitarity is respected up to higher orders as long as unstable particles are excluded from the final states. Renormalisation must nevertheless be treated carefully as not to spoil these properties of the bare theory. To this end, the standard on-shell scheme has been generalised in Ref.~\cite{Denner:2005fg} to the complex-mass case by requiring unit residues of propagators at complex poles and by taking into account the imaginary parts of the self-energies in the renormalisation constants. This necessitates, however, the calculation of the self-energies for complex squared momenta. This complication can be avoided by expanding the self-energies present in the renormalisation constants about real arguments such that one-loop accuracy is preserved.  In case of the  $W^\pm$ gauge boson and top-quark, extra constants must be added to the expanded self-energies in order not to spoil the one-loop accuracy of the results due to the presence of branch cuts introduced by infrared divergences. The simplified version of the complex renormalisation is described in detail in Refs.~\cite{Denner:2005fg,Denner:2019vbn}, where all renormalisation constants are provided as well. These renormalisation constants are implemented in \Recola and employed in our calculation of NLO QCD and EW corrections to the $pp \to t\bar{t}\gamma(\gamma)$ process in the di-lepton top-quark decay channel without any modifications.  

As already mentioned in Section \ref{sec:desc-lo}, the presence of final-state photons in the Born-level process requires special care in the renormalisation of the electromagnetic coupling constant $\alpha$. In particular, we split the total power of $\alpha^n$ into $\alpha_{G_\mu}^{n-n_\gamma}\alpha (0)^{n_\gamma}$ where $n_\gamma =1$ ($n_\gamma =2$) for the underlying Born-level process $pp\to t\bar{t}\gamma (\gamma)$. The powers of the electromagnetic coupling constant associated with final-state photons are renormalised in the on-shell scheme (the $\alpha(0)$ scheme) \cite{Denner:2005fg,Denner:2019vbn,Pagani:2021iwa}. The remaining powers of $\alpha$ are renormalised in the $G_{\mu}$ scheme, where the renormalized electromagnetic coupling $\alpha$ is derived from the Fermi constant $G_F$, measured in muon decays, and the masses of the $W$ and $Z$ bosons. The renormalisation in this mixed scheme is realised by first performing the complete renormalisation of all powers of $\alpha$ in the $G_{\mu}$ scheme. The renormalisation scheme is then changed for $\alpha (0)^{n_\gamma}$ by introducing a new counterterm which is given by
\begin{equation}
\label{eq_alpha_ren}
2\,n_{\gamma}\,{\rm Re}\left(\delta Z_{e}\big|_{\alpha (0)} - \delta Z_{e}\big|_{G_\mu}\right) d\sigma^{\rm LO}=n_{\gamma}\,{\rm Re}\left( \Delta r^{(1)} \right) d\sigma^{\rm LO},
\end{equation}
where $\Delta r^{(1)}$ are the NLO EW corrections to the muon decay \cite{Sirlin:1980nh,Denner:1991kt,Hollik:1988ii,Marciano:1980pb} and $\delta Z_{e}\big|_{\alpha (0)}$ as well as $\delta Z_{e}\big|_{G_\mu}$ are the renormalisation constants of the electric charge $e$ in the on-shell and $G_\mu$ scheme, respectively. We note that in Eq.~\eqref{eq_alpha_ren} and in general in the whole calculation, the electromagnetic coupling is always set to $\alpha=\alpha_{G_\mu}$ and the final result is then rescaled by $(\alpha (0)/\alpha_{G_\mu})^{n_\gamma}$. This also implies that the relative EW corrections are always calculated with $\alpha=\alpha_{G_\mu}$. In order to test the consistency of the calculation in this mixed approach we have explicitly checked the cancellation of all IR singularities in the sum of virtual corrections and integrated dipoles for a few phase-space points for all partonic subprocesses.

%
\section{LHC Setup for numerical predictions}
\label{sec:setup}
%

The calculations of complete NLO corrections for $pp\to t\bar{t}\gamma$ and $pp\to t\bar{t}\gamma\gamma$ in the di-lepton top-quark decay channel are performed for proton-proton collisions at the LHC with a center of mass energy of ${\sqrt{s} = 13}$ TeV. When both higher-order QCD and EW corrections as well as photon-initiated subprocesses are considered, the PDFs with QED effects and the photon content of the proton are essential. To this end, we use the NLO NNPDF3.1luxQED PDF set \cite{Manohar:2016nzj,NNPDF:2017mvq,Manohar:2017eqh,Bertone:2017bme} with $\alpha_s(m_Z)=0.118$, where photons are properly taken into account in the PDF evolution, at LO and NLO. The PDF set as well as the running of the strong coupling constant with two-loop accuracy is obtained using the LHAPDF interface \cite{Buckley:2014ana}. As already explained in the last section,  for $\alpha$ a mixed scheme is considered. In particular,  $\alpha$ is first calculated in the $G_\mu$-scheme according to
\begin{equation}
	\alpha_{G_\mu} =\frac{\sqrt{2}}{\pi} \,G_\mu \, m_W^2\,\left(1-\frac{m_W^2}{m_Z^2}\right)\,,
	~~~~~~~~~~~~~~~~~~~~~
	G_{ \mu}=1.1663787 \cdot 10^{-5} \textrm{ GeV}^{-2}\,,
\end{equation}
with $m_{W}= 80.379$ GeV and $m_{Z}=91.1876$ GeV. However, $\alpha$ associated with Born-level final-state photons is treated in the on-shell scheme, where we use $\alpha^{-1}= \alpha^{-1}(0) =137.035999084$ \cite{Workman:2022ynf} as the input parameter. Additional photon radiation from the real emission part of the NLO calculation is evaluated with $\alpha_{G_\mu}$. We use the following input values for the masses and widths of the unstable particles
\begin{equation}
\begin{array}{lll}
m_{t}=172.5 ~{\rm GeV}\,,&\quad\quad\quad\quad &  \Gamma_{W} = 2.0972 ~{\rm GeV}\,, 
\vspace{0.2cm}\\
 m_{H}=125 ~{\rm GeV}\,,&\quad\quad\quad\quad & \Gamma_{H}=4.07\cdot 10^{-3}~{\rm GeV}\,.
\end{array}
\end{equation}
The width of the $Z$ boson is set to zero to avoid artificial higher-order terms that would otherwise appear due to the complex-mass scheme, since the width of the $W$ boson is assumed to be zero everywhere except in the propagators, as explained in Ref.~\cite{Denner:2016jyo}. All other particles are assumed to be massless. We use the same top-quark width at LO and NLO. Taking into account NLO QCD and EW corrections, the latter width is given by
\begin{equation}
\Gamma_{t}^{\rm NLO} = 1.3735727  ~{\rm GeV},
\end{equation}
where we adapt the conventions given in Ref.~\cite{Denner:2012yc}, while the NLO QCD corrections are obtained from  Ref.~\cite{Jezabek:1988iv} for $\mu_R=m_t$. We note here that, contrary to what is usually done when calculating NLO QCD corrections, in this study we use the same PDF set as well as top-quark width for both LO and NLO predictions. This should help us to directly examine the relative magnitude of the subleading contributions, while simultaneously separating out the effects coming from the use of different settings at LO and NLO. The NLO EW corrections are computed numerically with the help of \helacdipoles and \Recola. As already indicated in Eq.~\eqref{eq:nlo_tot}, in this scheme, the LO and NLO contributions can simply be added to give the complete result. 

The event selection and scale choice are based on our previous work on NLO QCD corrections to the $pp \to t\bar{t}\gamma\gamma$ process \cite{Stremmer:2023kcd}. All final-state partons with pseudo-rapidity $|\eta|<5$ are clustered  into jets with separation $R=0.4$ in the rapidity azimuthal-angle plane via the IR-safe  $anti$-$k_T$ jet algorithm \cite{Cacciari:2008gp}. We require two oppositely charged leptons, at least two $b$-jets and at least one (two) photon(s) for $t\bar{t}\gamma(\gamma)$, respectively. The IR safety of the $t\bar{t}\gamma(\gamma)$ cross section involving jets and prompt photon(s) is ensured by using the smooth photon isolation prescription as described in Ref.~\cite{Frixione:1998jh}. Therefore, the event is rejected unless the following condition is fulfilled before the jet algorithm is applied
\begin{equation}
 \label{eq_iso}
\sum_{i} E_{T\,i}  \, \Theta(R - R_{\gamma i})  \le \epsilon_\gamma \, E_{T\,\gamma} \left(
\frac{1-\cos(R)}{1-\cos(R_{\gamma j})}
\right)^n \,,
\end{equation}
for all $R\le R_{\gamma j}$ with $R_{\gamma j}=0.4$ and $\epsilon_{\gamma}=n=1$. The transverse energy of the parton $i$/photon is denoted by $E_{T\,i}$/$E_{T\,\gamma}$ and $R_{\gamma i}$ is given by 
\begin{equation}
R_{\gamma i}=\sqrt{(y_\gamma-y_i)^2+(\phi_\gamma-\phi_i)^2}\,.
\end{equation}
In subleading NLO corrections we also encounter partonic processes with an additional photon in the final state with respect to the Born-level subprocesses. In this case, the smooth photon isolation prescription can be extended to additionally require the isolation of photons with charged leptons and photons, as suggested in Ref.~\cite{Pagani:2021iwa}. In this approach, after the smooth photon isolation, charged leptons and partons are recombined with only non-isolated photons based on a jet clustering algorithm. In general, the use of the smooth photon isolation prescription is not required in this case. Alternatively, it is sufficient to perform the photon recombination directly with charged leptons and partons, where additionally the following recombination rule $\gamma + \gamma\to\gamma$ is allowed. Similar schemes have already been applied in associated photon production processes, see e.g. \ Refs.~\cite{Denner:2014bna,Denner:2015fca}. We have used both approaches for the calculation of  higher-order corrections to the $pp\to t\bar{t}\gamma$ process, where the photon recombination has  also been performed with the $anti$-$k_T$ jet algorithm with the radius parameter $R=0.4$. We have found no phenomenologically relevant differences even at the differential cross-section level. Therefore, we remain with the second approach, not using the smooth isolation prescription in these contributions. After the jet algorithm/photon recombination, the prompt photons are defined with the following conditions
\begin{equation}
\begin{array}{lll}
p_{T,\,\gamma}>25 ~{\rm GeV}\,,  
&\quad \quad\quad\quad\quad |y_\gamma|<2.5 \,, 
 &\quad \quad\quad \quad \quad
\Delta R_{\gamma\gamma}>0.4\,.
\end{array}
\end{equation}
In order to ensure well-observable isolated $b$-jets and charged leptons in the central-rapidity region, we require
\begin{equation}
\begin{array}{lll}
p_{T,\,b}>25 ~{\rm GeV}\,,  
&\quad \quad\quad\quad\quad |y_b|<2.5 \,, 
 &\quad \quad\quad \quad \quad
\Delta R_{bb}>0.4\,,
\\[0.2cm]
 p_{T,\,\ell}>25 ~{\rm GeV}\,,    
 &\quad \quad \quad \quad\quad|y_\ell|<2.5\,,&
\quad \quad \quad \quad \quad
\Delta R_{\ell
 \ell} > 0.4\,.
\end{array}
\end{equation}
In addition,  charged leptons and photons need to  be well separated from any $b$-jet in the rapidity-azimuthal angle plane as well as from each other
\begin{equation}
\begin{array}{lll}
\Delta R_{l\gamma}>0.4\,,  
&\quad \quad\quad\quad\quad \Delta R_{lb}>0.4 \,, 
 &\quad \quad\quad \quad \quad
\Delta R_{b\gamma}>0.4\,.
\end{array}
\end{equation}
Finally, there are no restrictions on the kinematics of the extra light jet/photon (if resolved by the jet algorithm) and the missing transverse momentum, $p_T^{miss} = |\vec{p}_{T,\, \nu_\ell} + \vec{p}_{T,\, \bar{\nu}_\ell}|$. Based on the findings in Ref.~\cite{Stremmer:2023kcd} we set the factorisation and renormalisation scale to a common dynamical scale  setting, $\mu_0$, given by
\begin{equation}
\mu_R=\mu_F=\mu_0=\frac{E_T}{4} \,, 
\end{equation}
with
\begin{equation}
\label{eq_scale}
E_T=\sqrt{m^2_{t}+p_{T, \,t}^2}+\sqrt{m^2_{t}+p_{T,\, \bar{t}}^2 }  + \sum_{i=1}^{n_\gamma} p_{T,\,\gamma_i}\,,
\end{equation} 
where $p_{T,\,t}$ and $p_{T,\,\bar{t}}$ are the transverse momenta of the on-shell top quarks.  In addition, for dimensionful cross-section distributions constructed from the photon kinematics, we present results for the fixed scale setting, $\mu_0=m_t$. Theoretical uncertainties due to missing higher orders are estimated by a 7-point scale variation, where the factorisation and renormalisation scales are independently varied in the range
\begin{equation}
\frac{1}{2} \, \mu_0  \le \mu_R\,,\mu_F \le  2 \,  \mu_0\,, \quad \quad 
\quad \quad \quad \quad \quad \quad \frac{1}{2}  \le
\frac{\mu_R}{\mu_F} \le  2 \,,
\end{equation}
which leads to the following pairs
\begin{equation}
\label{scan}
\left(\frac{\mu_R}{\mu_0}\,,\frac{\mu_F}{\mu_0}\right) = \Big\{
\left(2,1\right),\left(0.5,1  
\right),\left(1,2\right), (1,1), (1,0.5), (2,2),(0.5,0.5)
\Big\} \,.
\end{equation}
By searching for the minimum and maximum of the resulting cross sections we obtain the displayed uncertainty bands. We use the 7-point scale variation both at the integrated and differential cross-section level.

%
\section{Top-quark pair production with one isolated photon}
\label{sec:tta}
%

%
\subsection{Integrated fiducial cross sections}
\label{sec:tta-int}
%

Now we turn to the discussion of the numerical results. We begin with our findings for the $pp\to t\bar{t}\gamma$ process at the integrated fiducial cross-section level. In particular, in Table \ref{tab:tta} we present the results for \LOfull, \NLOfull, \NLOqcd and \NLOprd, as defined in the last section, with the corresponding theoretical uncertainties obtained by scale variation. In addition, we display the decomposition of the LO and NLO results into the individual ${\rm LO}_i$ and ${\rm NLO}_i$ contributions calculated at the different orders in $\alpha_s$ and $\alpha$. Finally, the relative size of all results with respect to the dominant LO contribution, \LOone, is given in the last column.
\begin{table*}[t!]
    \centering
    \renewcommand{\arraystretch}{1.2}
    \begin{tabular}{ll@{\hskip 10mm}l@{\hskip 10mm}l@{\hskip 10mm}}
        \hline
        \noalign{\smallskip}
         &&$\sigma_{i}$ [fb] & Ratio to ${\rm LO}_1$  \\
        \noalign{\smallskip}\midrule[0.5mm]\noalign{\smallskip}
        ${\rm LO}_1$&$\mathcal{O}(\alpha_s^2\alpha^5)$& $ 55.604(8)^{+31.4\%}_{-22.3\%} $ & $ 1.00 $ \\
        ${\rm LO}_2$&$\mathcal{O}(\alpha_s^1\alpha^6)$& $ 0.18775(5)^{+20.1\%}_{-15.4\%} $ & $ +0.34\% $ \\
        ${\rm LO}_3$&$\mathcal{O}(\alpha_s^0\alpha^7)$& $ 0.26970(4)^{+14.3\%}_{-16.9\%} $ & $ +0.49\% $ \\
        \noalign{\smallskip}\hline\noalign{\smallskip}
        ${\rm NLO}_1$&$\mathcal{O}(\alpha_s^3\alpha^5)$& $ +3.44(5) $ & $ +6.19\% $\\
        ${\rm NLO}_2$&$\mathcal{O}(\alpha_s^2\alpha^6)$& $ -0.1553(9) $ & $ -0.28\% $\\
        ${\rm NLO}_3$&$\mathcal{O}(\alpha_s^1\alpha^7)$& $ +0.2339(3) $ & $ +0.42\% $\\
        ${\rm NLO}_4$&$\mathcal{O}(\alpha_s^0\alpha^8)$& $ +0.001595(8) $ & $ +0.003\% $\\
        \noalign{\smallskip}\hline\noalign{\smallskip}
        ${\rm LO}$&& $ 56.061(8)^{+31.2\%}_{-22.1\%} $ & $ 1.0082 $ \\
        ${\rm NLO}_{\rm QCD}$&& $ 59.05(5)^{+1.6\%}_{-5.9\%} $ & $ 1.0620 $ \\
        ${\rm NLO}_{\rm prd}$&& $ 59.08(5)^{+1.5\%}_{-5.9\%} $ & $ 1.0626 $ \\
        ${\rm NLO}$&& $ 59.59(5)^{+1.6\%}_{-5.9\%} $ & $ 1.0717 $ \\
        \noalign{\smallskip}\hline\noalign{\smallskip}
    \end{tabular}
    \caption{\label{tab:tta} \it Integrated fiducial cross section for the $pp\to t\bar{t}\gamma +X$ process in the di-lepton top-quark decay channel at the LHC with $\sqrt{s}=13$ TeV. Results are shown for \LOfull, \NLOfull, \NLOqcd and \NLOprd with the corresponding scale uncertainties. Additionally, all ${\rm LO}_i$ and ${\rm NLO}_i$ contributions and the ratio with respect to \LOone are displayed. MC integration errors are given in parentheses. Results are provided for $\mu_0 = E_T/4$ and the NLO NNPDF3.1luxQED PDF set.}
\end{table*}
We find that both subleading LO contributions, \LOtwo and \LOthree, are below $0.5\%$ of \LOone. In addition,  \LOthree is slightly larger than \LOtwo due to cancellations in \LOtwo between the $g\gamma$ and $b\bar{b}$  subprocesses.  It turns out that the latter channel gives a negative result that can be attributed to the interference between the different orders in $\alpha_s$ and $\alpha$. Although, the $b\bar{b}$ initial state is PDF suppressed with respect to the $q\bar{q}$ one, it  provides the dominant contribution to \LOthree, and is more than three times larger than the $q\bar{q}$ channel, due to the presence of additional $t$-channel Feynman diagrams that do not appear in the $q\bar{q}$ subprocess. The $\gamma\gamma$ initiated contribution is smaller than the MC integration error of the \LOfull result and thus completely negligible.  Consequently, at the integrated fiducial cross-section level, the subleading LO contributions are phenomenologically insignificant, especially when compared to the scale uncertainties of the complete LO result (or the dominant \LOone contribution for that matter). The theoretical errors of both LO and \LOone are of similar size and amount to $31\%$. 

As expected, the dominant NLO corrections originate from the NLO QCD corrections (\NLOone) to \LOone and lead to an increase of the integrated fiducial cross section by about $6.2\%$. All subleading NLO contributions are again found to be less than $0.5\%$ of \LOone.  In more detail,  \NLOtwo and \NLOthree are of the same order of magnitude and \NLOfour is completely negligible compared to the  NLO scale uncertainties and even MC integration errors. We note that in all the contributions we used the same top-quark width ($\Gamma_t^{\rm NLO, QCD+EW} = 1.3735727$ GeV).  In particular, if the LO top-quark width ($\Gamma_t^{\rm LO} = 1.4806842$ GeV) has been used instead for \LOone and the NLO QCD one ($\Gamma_{t}^{\rm NLO, QCD}= 1.3535983$ GeV) for \NLOqcd, as it is often done in higher-order QCD calculations, we would obtain NLO QCD corrections of about $27\%$. Furthermore, we note that the sum of all subleading LO and NLO contributions increases the \NLOqcd result by about $1\%$ and therefore plays only a minor role at the integrated fiducial cross-section level compared to the \NLOqcd scale uncertainties that are of the order of $6\%$. The scale uncertainties themselves remain unchanged by the inclusion of the subleading higher-order contributions, and are thus completely  driven by the ${\rm NLO}_{\rm QCD}$ part.  Finally, we can observe that  ${\rm NLO}_{\rm prd}$ and ${\rm NLO}_{\rm QCD}$ agree within the MC errors,  which results from the cancellation of subleading LO contributions ($0.8\%$) and the subleading NLO corrections to the production of $pp\to t\bar{t}\gamma$ ($-0.7\%$). In addition, the difference between \NLOfull and \NLOprd, and thus the subleading NLO corrections involving radiative top-quark decays (including photon radiation and/or QCD/EW corrections), is of similar (absolute) size as the other two effects $(0.9\%)$.

%
\subsection{Differential fiducial cross sections}
\label{sec:tta-diff}
%

\begin{figure}[t!]
    \begin{center}
	\includegraphics[width=0.49\textwidth]{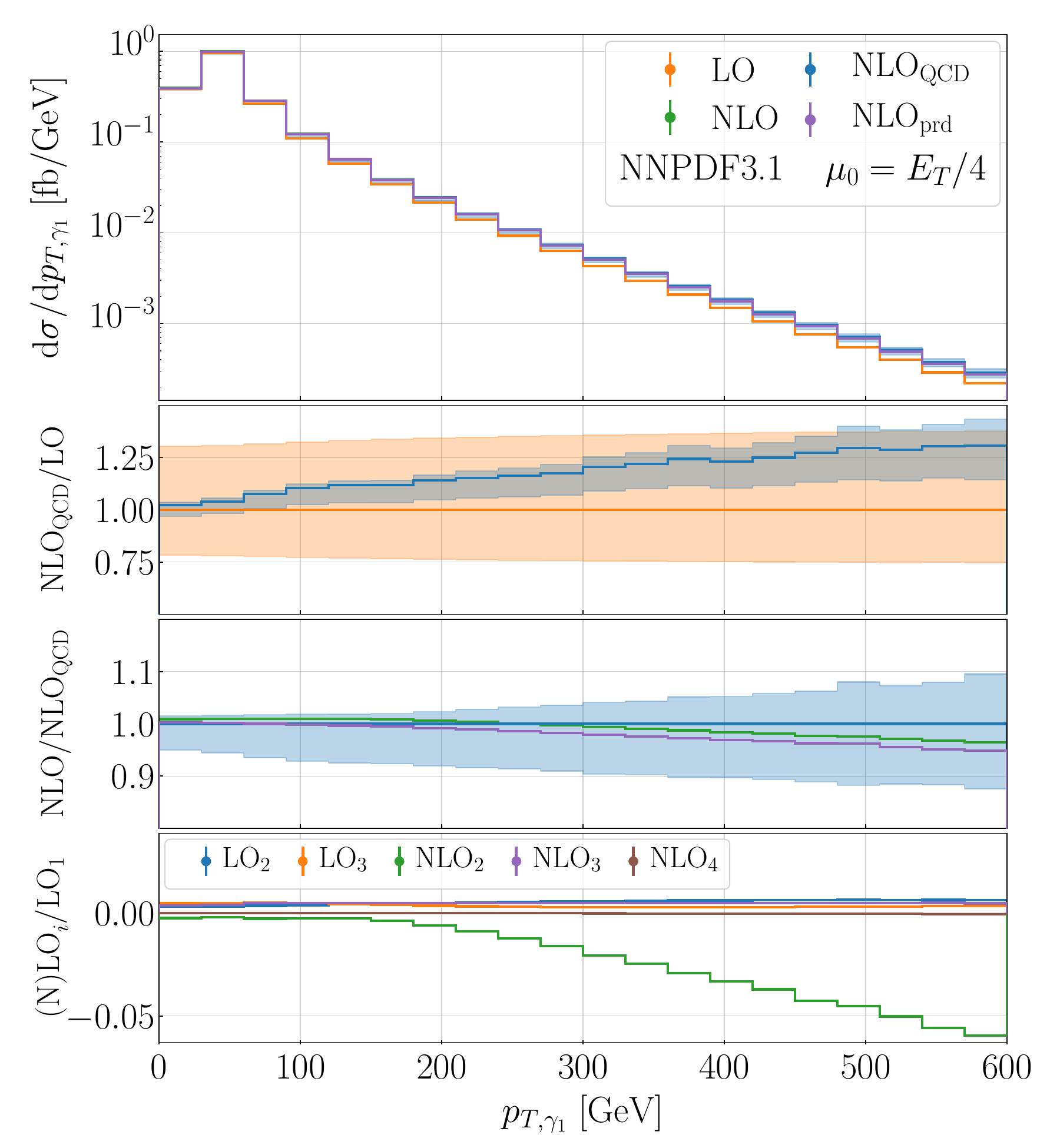}
	\includegraphics[width=0.49\textwidth]{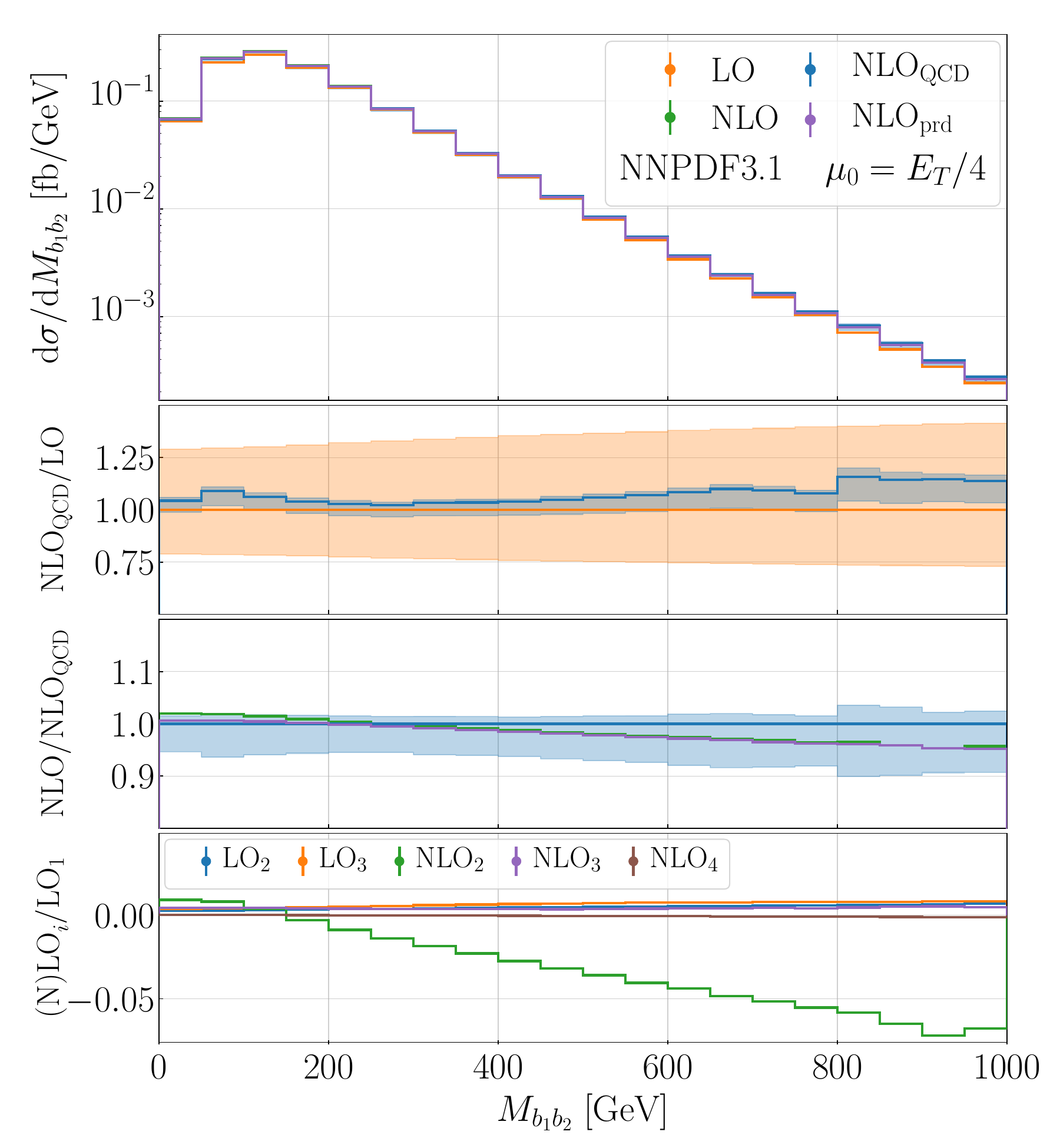}
	\includegraphics[width=0.49\textwidth]{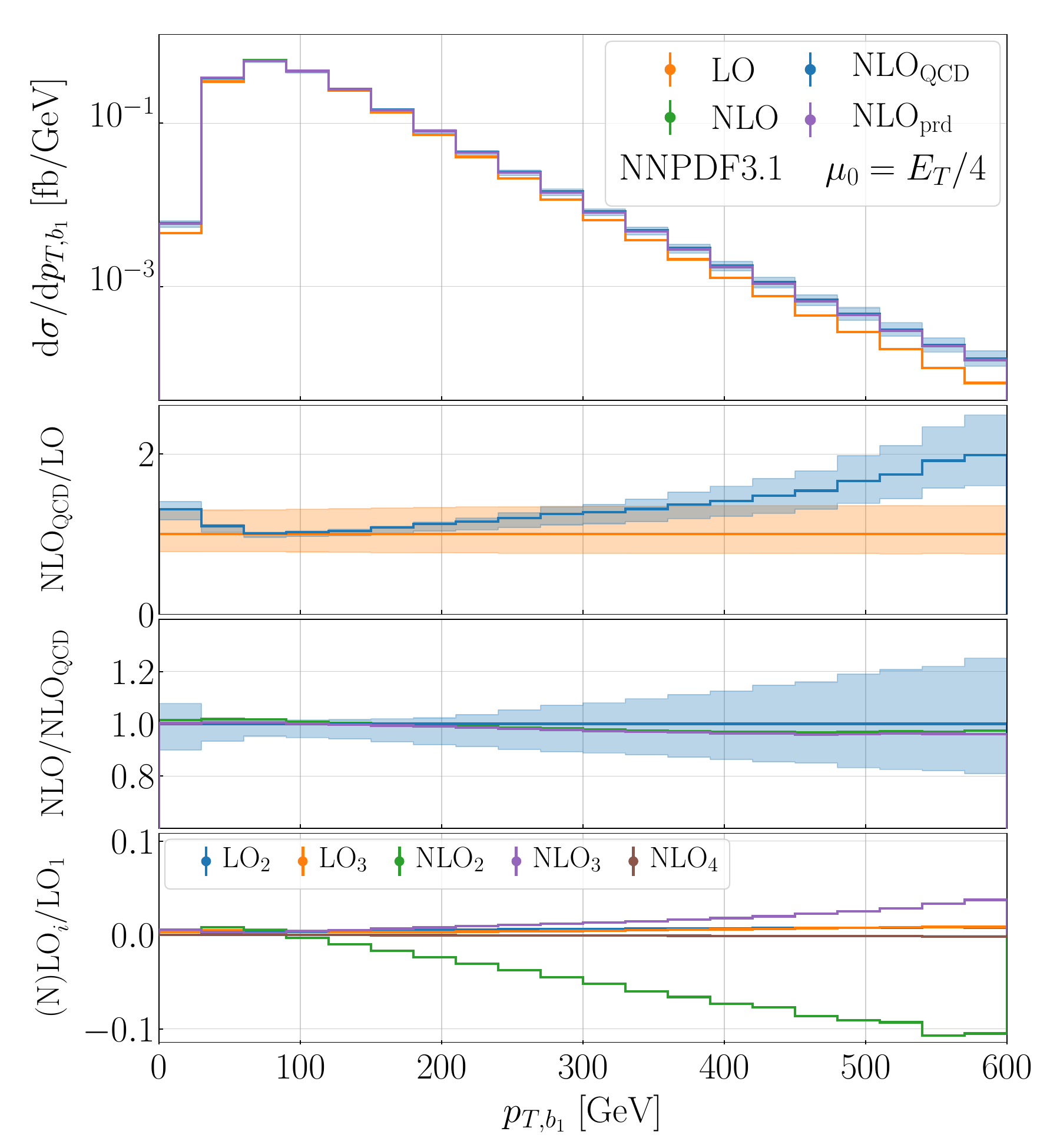}
	\includegraphics[width=0.49\textwidth]{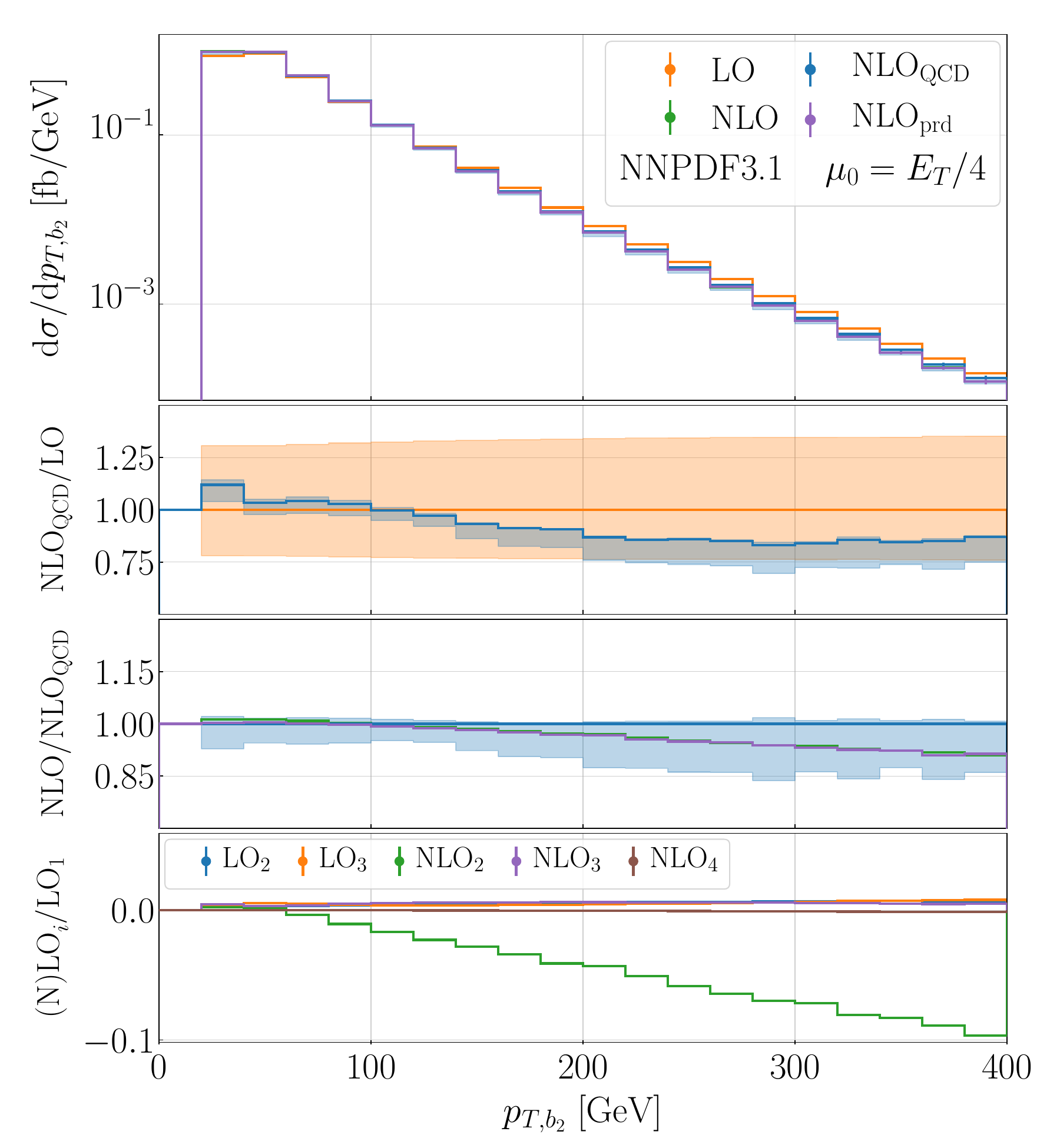}
    \end{center}
    \caption{\label{fig-tta:nlo1} \it Differential cross-section distributions for the observables $p_{T,\gamma_1}$, $M_{b_1b_2}$, $p_{T,b_1}$ and $p_{T,b_2}$ for the $pp\to t\bar{t} \gamma +X$ process in the di-lepton top-quark decay channel at the LHC with $\sqrt{s}=13$ TeV. The upper panels present absolute predictions for \LOfull, \NLOfull, \NLOqcd and \NLOprd  together with the \NLOqcd  uncertainty bands resulting from scale variations. The middle panels show the ratio of the \NLOqcd results to \LOfull along with their relative scale uncertainties as well as the ratio of the \NLOfull and \NLOprd results with respect to \NLOqcd. In the later case also given are the \NLOqcd  uncertainty bands. Finally, the lower panels display the relative size of all subleading ${\rm LO}_i$ and ${\rm NLO}_i$ contributions compared to \LOone. Results are provided for $\mu_F=\mu_R=\mu_0 = E_T/4$ and the NLO NNPDF3.1luxQED PDF set.}
\end{figure}
\begin{figure}[t!]
    \begin{center}
	\includegraphics[width=0.49\textwidth]{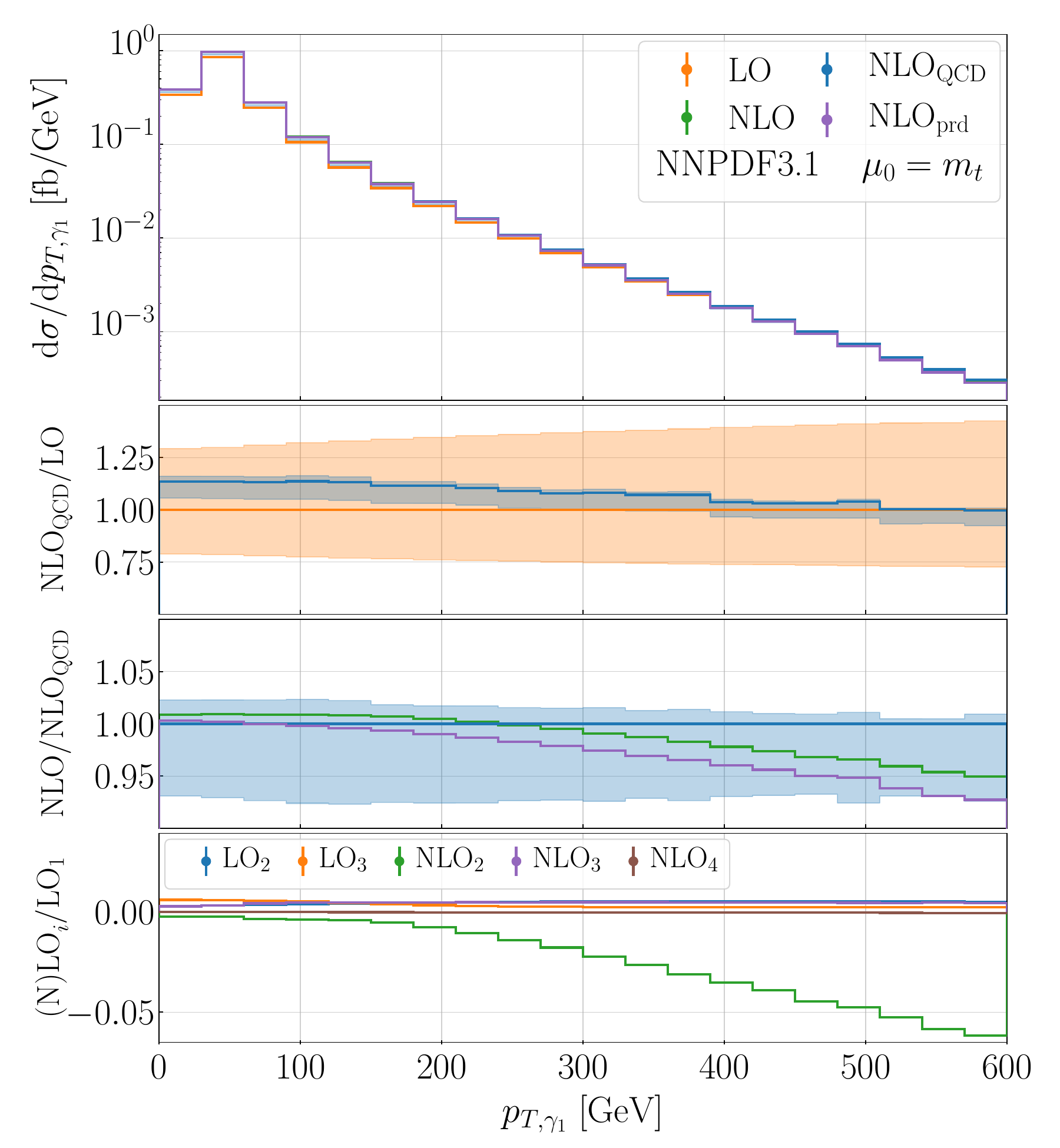}
    \end{center}
    \caption{\label{fig-tta:nlo1_mt} \it Same as Figure \ref{fig-tta:nlo1} but for the observable $p_{T,\gamma_1}$ with  $\mu_R=\mu_F=\mu_0=m_t$.}
\end{figure}
\begin{figure}[t!]
    \begin{center}
	\includegraphics[width=0.49\textwidth]{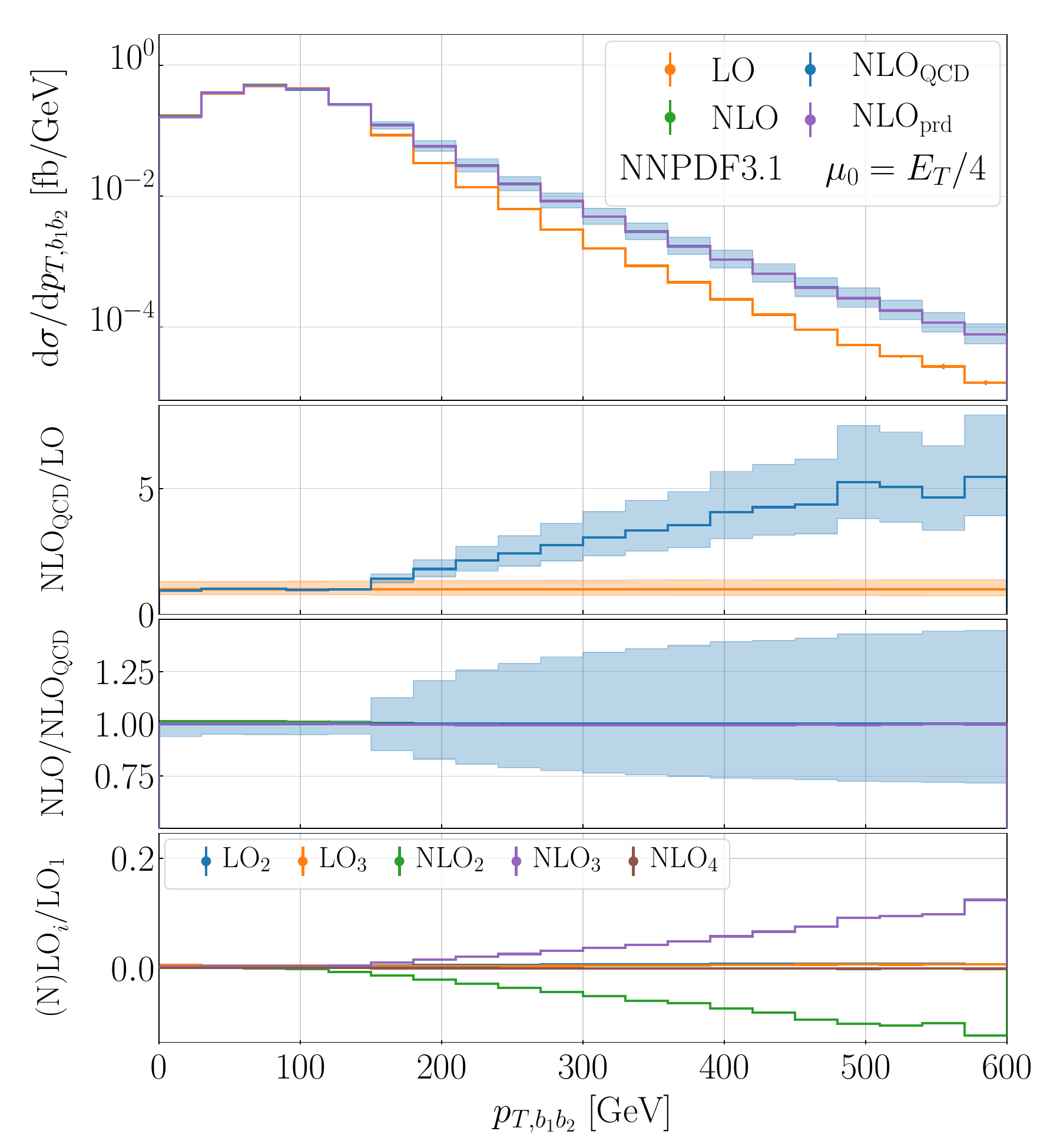}
	\includegraphics[width=0.49\textwidth]{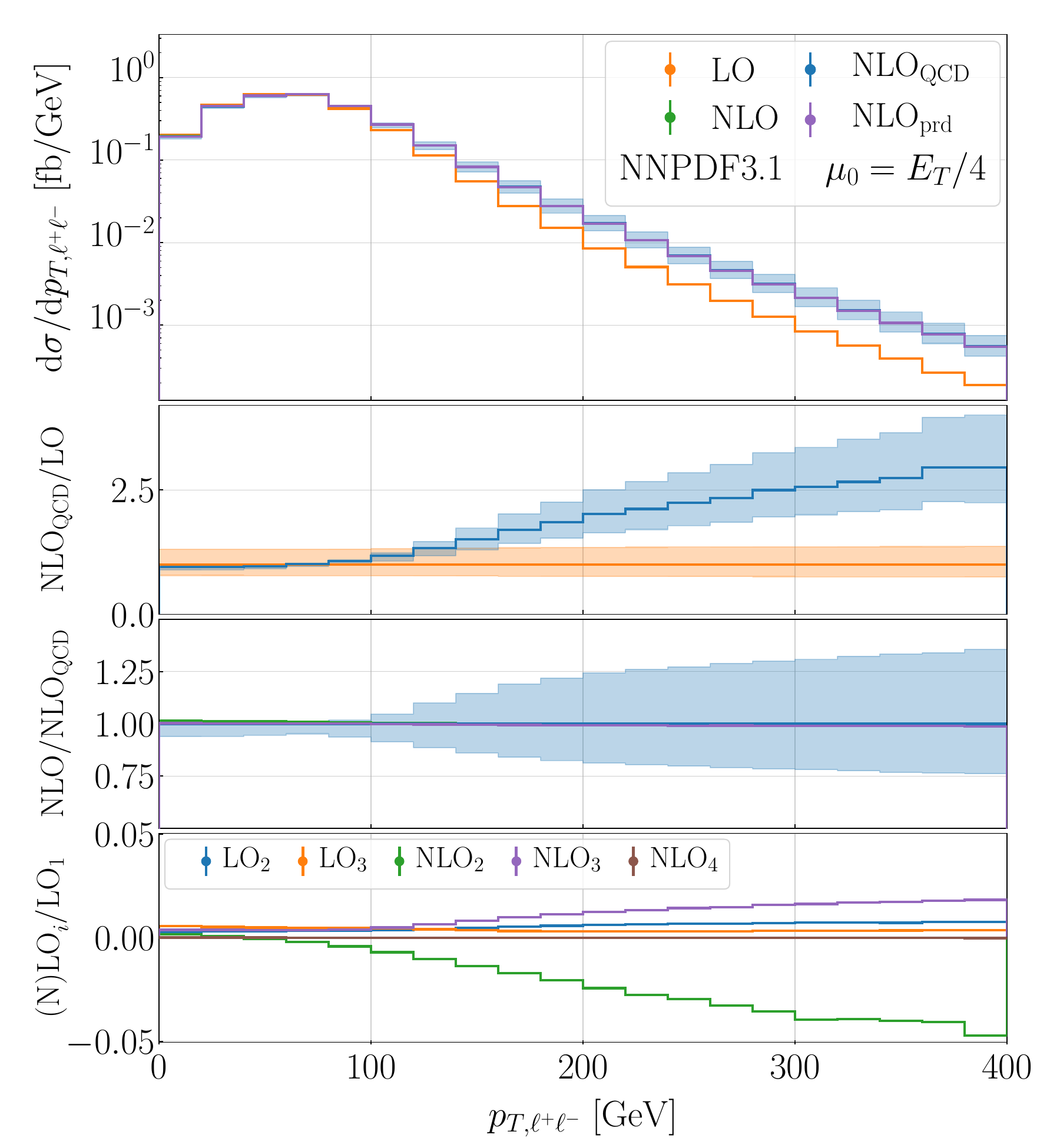}
	\includegraphics[width=0.49\textwidth]{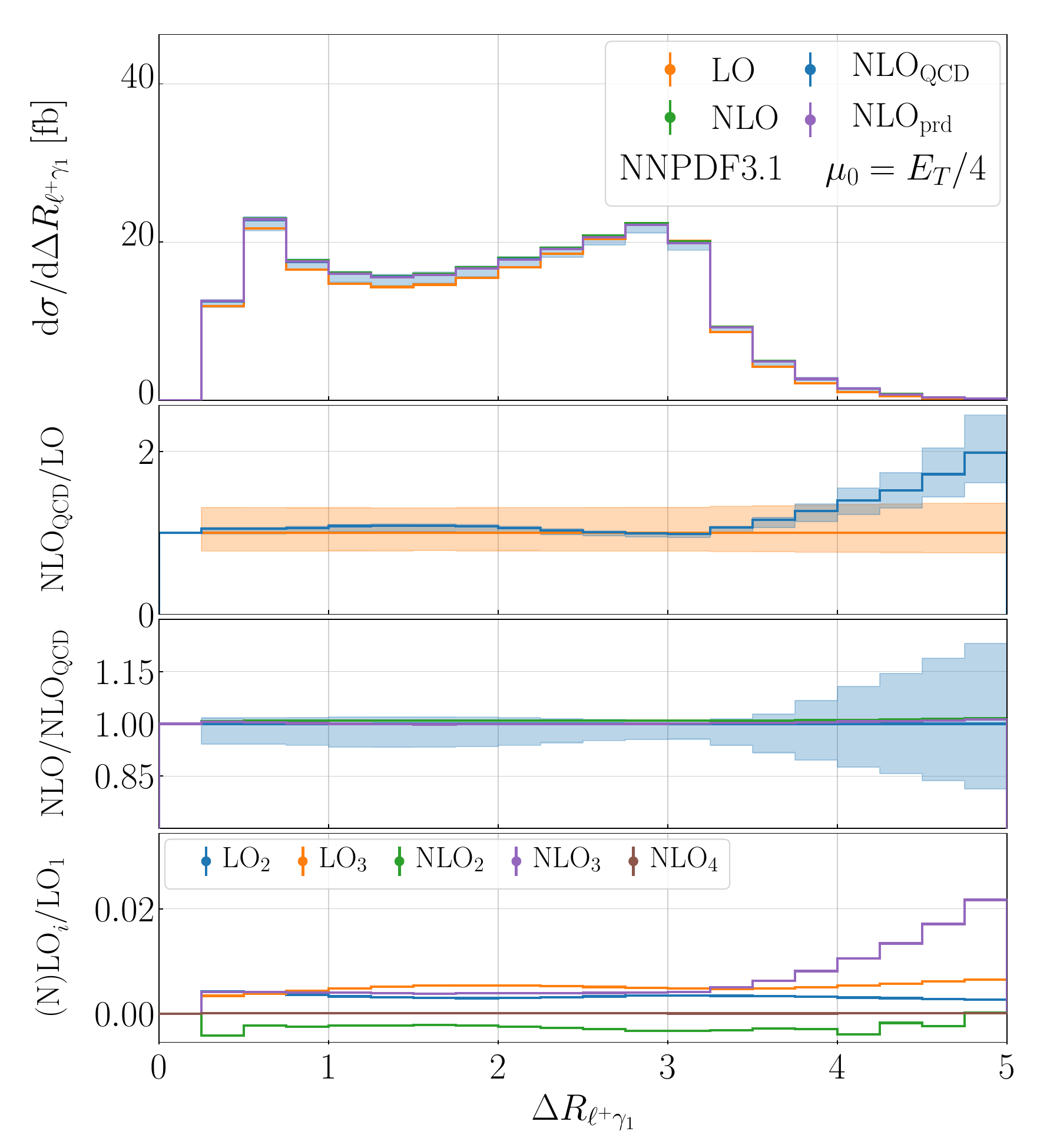}
	\includegraphics[width=0.49\textwidth]{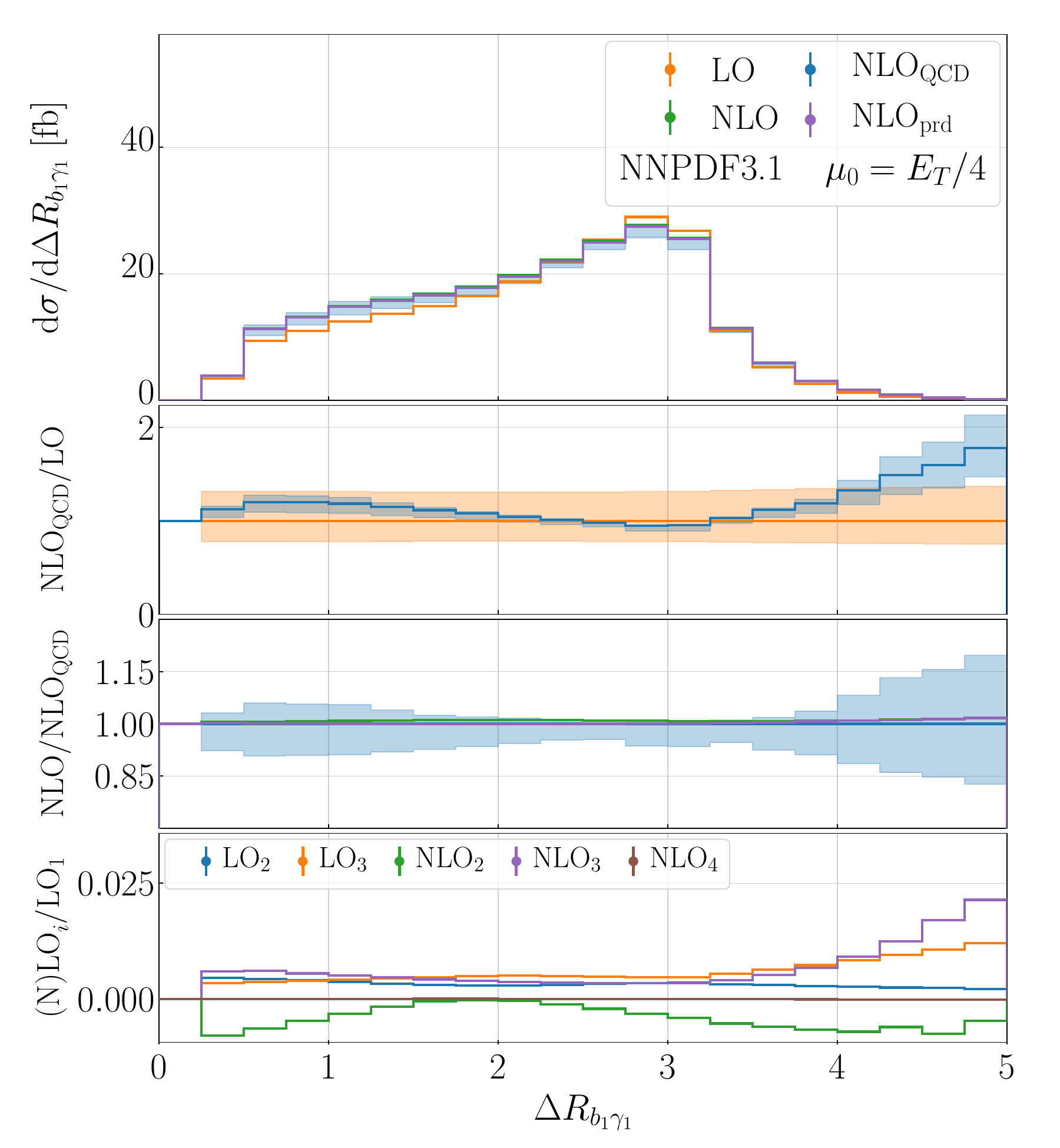}
    \end{center}
    \caption{\label{fig-tta:nlo2} \it Same as Figure \ref{fig-tta:nlo1} but for the observables $p_{T,b_1b_2}$, $p_{T,\ell^+\ell^-}$, $\Delta R_{\ell^+\gamma_1}$ and $\Delta R_{b_1\gamma_1}$. }
\end{figure}

While at the integrated fiducial cross-section level the effects of the subleading LO and NLO contributions are small, it is expected that these effects might be enhanced for particular observables in certain phase-space regions. To this end, in Figure \ref{fig-tta:nlo1} we present the differential cross-section distributions for the transverse momentum of the hardest photon, $p_{T,\gamma_1}$, the invariant mass of the two hardest $b$-jets, $M_{b_1b_2}$, as well as their transverse momenta, denoted as $p_{T,b_1}$ and $p_{T,b_2}$, respectively.  In the upper panels we provide the absolute predictions for \LOfull, \NLOfull, \NLOqcd and \NLOprd together with the \NLOqcd uncertainty bands resulting from scale variations. The middle panels show the ratio of the \NLOqcd results to \LOfull along with their relative scale uncertainties as well as the ratio of the \NLOfull and \NLOprd results to \NLOqcd. In the latter case the \NLOqcd theoretical uncertainties are also displayed. Finally, the lower panels present the relative size of all subleading ${\rm LO}_i$ and ${\rm NLO}_i$ contributions with respect to \LOone. 

For $p_{T,\gamma_1}$ we find that the largest NLO corrections arise from \NLOone and lead in the tail of the distribution to an increase of the \LOone cross section by about $30\%$. In these phase-space regions the \NLOqcd scale uncertainties increase to about $12\%$. The inclusion of all subleading LO and NLO contributions reduces the \NLOqcd prediction by up to $4\%$ due to EW Sudakov logarithms induced by \NLOtwo. All other subleading contributions remain below $1\%$. The \NLOprd contribution, where the subleading NLO corrections are only included in the production stage of $t\bar{t}\gamma$, underestimates the full calculation by about $2\%$. The overall picture remains the same for the $M_{b_1b_2}$ observable. In particular, we again find that \NLOtwo is the dominant subleading contribution, leading to a reduction in the \NLOqcd  prediction of about $5\%$. This reduction is about half the size of the corresponding scale uncertainties. On the other hand, \NLOprd recovers the full calculation correctly in the tails and only small differences up to $1.5\%$ are found at the beginning of the spectrum. In the case of the transverse momenta of the hardest and second hardest $b$-jet, $p_{T,b_1}$ and $p_{T,b_2}$, the relative size of \NLOtwo with respect to \LOone is increased to $10\%$. On the contrary, the impact of all subleading contributions on the final NLO result seems to be very different. Indeed, we find for $p_{T,b_1}$ that \NLOthree is enhanced in the tail and amounts to $3\%-4\%$, thus partially cancelling the EW Sudakov logarithms in \NLOtwo. Moreover, we find large NLO QCD corrections to \LOone of up to $95\%$, which further suppress the importance of the subleading contributions. However, both effects, the enhancement of \NLOthree and the large \NLOone contribution, have the same origin and arise from large NLO QCD real corrections to \LOone and \LOthree, induced by hard jet recoil against the $t\bar{t}$ system. This can be seen by inspecting the size of the \NLOqcd scale uncertainties, which increase continuously towards the tail of the distribution and become closer in size to those at LO. Thus, the relative size of the subleading corrections is highly dependent on the fiducial phase space and might be enhanced if a more exclusive event selection and/or jet vetos are applied. On the other hand, for $p_{T,b_2}$ the subleading contributions have a more significant impact on its spectrum, reducing the \NLOqcd prediction by up to $10\%$. The scale uncertainties of \NLOqcd are about $15\%$ and therefore comparable in size. Moreover, we find that the larger size of the subleading contributions can also affect the scale uncertainties, which in this case increase to about $20\%$ for the complete \NLOfull case. Finally, we can observe that, for both $b$-jet transverse momentum spectra the \NLOprd predictions fully recover the complete NLO result.

Following the findings of our previous work on the calculation of the NLO QCD corrections to the $pp\to t\bar{t}\gamma\gamma$ process in the di-lepton top-quark decay channel \cite{Stremmer:2023kcd}, where a traditional fixed scale setting, $\mu_R=\mu_F=\mu_0=m_t$, led to a reduction of NLO QCD corrections and scale uncertainties for most photonic observables, in the following we reexamine the $p_{T,\gamma_1}$ observable also for this scale choice. To this end, in Figure \ref{fig-tta:nlo1_mt} we present $p_{T,\gamma_1}$ afresh for $\mu_0=m_t$. We find that for this fixed scale choice the \NLOone contribution is reduced from $25\%$ to $13\%$. Moreover, while for the default scale setting, $\mu_0=E_T/4$, the largest NLO QCD corrections are present in the tail, for $\mu_0=m_t$ the \LOfull and \NLOqcd predictions coincide in this phase-space region. Furthermore, the differences of up to $13\%$ are found at the beginning of the spectrum. The relative size of the subleading LO and NLO contributions is basically unchanged and leads to a reduction of the complete calculation of about $5\%$ compared to $4\%$ for $\mu_0=E_T/4$. In addition, the scale uncertainties are slightly reduced to $7\%-9\%$ compared to $10\%-11\%$. Thus, for the scale setting $\mu_0=m_t$, the subleading contributions become more important, since the subleading effects and the scale uncertainties are of similar size.

We continue the discussion and present in Figure \ref{fig-tta:nlo2} other differential cross-section distributions. In detail, we display the transverse momentum of the $b_1b_2$ and $\ell^+\ell^-$ systems, denoted as $p_{T,b_1b_2}$ and $p_{T,\ell^+\ell^-}$, respectively. Furthermore, we show the angular separation between $\gamma_1$ and $\ell^+$ as well as $\gamma_1$ and $b_1$, denoted as $\Delta R_{\ell^+\gamma_1}$ and $\Delta R_{b_1\gamma_1}$, respectively. The two dimensionful observables, $p_{T,b_1b_2}$ and $p_{T,\ell^+\ell^-}$, are affected by huge NLO QCD corrections to \LOone of about $450\%$ and $200\%$. These corrections arise from a kinematical suppression at LO \cite{Denner:2012yc,Denner:2015yca,Bevilacqua:2019cvp,Czakon:2020qbd,Stremmer:2021bnk}. Indeed, at the LO level the two top quarks are predominantly produced in the back-to-back configuration, so that the transverse momentum of both $b$-jets or leptons may separately be large, but they largely cancel each other out in the combined system. At NLO, however, this suppression is weakened due to hard jet recoil against the $t\bar{t}$ system. This leads, on the one hand, to huge NLO QCD corrections to \LOone, and on the other hand, also to an enhancement of the NLO QCD corrections with respect to \LOthree as part of \NLOthree. In particular, for $p_{T,b_1b_2}$ we find that \NLOthree is about $10\%$ of \LOone. The \NLOtwo contribution is of similar magnitude but has the opposite sign. Consequently, \NLOthree and \NLOtwo cancel each other out, and the \NLOqcd and \NLOfull predictions coincide. Since the two NLO subleading contributions are of different origins, this cancellation is to a large extend accidental and highly dependent on the event selection and for example jet vetos. Especially, vetoing undesired additional jets would drastically reduce the size of \NLOone and \NLOthree. Also for $p_{T,\ell^+\ell^-}$ we find a similar but less pronounced cancellation of \NLOtwo ($-5\%$) and \NLOthree ($2\%$). The \NLOprd approximation is again able to fully recover the complete NLO calculation. 

For angular distributions such as regular rapidity distributions, $\Delta R$ separations or (azimuthal) opening angles between two final states, we find no significant effects from any of the subleading LO or NLO contributions. Indeed, they are individually in the range of $0.5\%-1.0\%$ when compared to \LOone. Only for the $\Delta R$ separation between a photon and a $b$-jet or charged lepton, such as $\Delta R_{\ell^+\gamma_1}$ and $\Delta R_{b_1\gamma_1}$, we find a small enhancement for the \NLOthree contribution when comparing to \LOone. This enhancement, of about $2\%$, is visible for large values of the $\Delta R$ separation. However, these phase-space regions are generally not only less populated but also characterized by substantial NLO scale uncertainties up to $15\%$. Such theoretical uncertainties are induced by large NLO QCD corrections to \LOone. Thus, this small enhancement of \NLOthree is still negligible.

In summary, subleading LO and NLO contributions are generally only important in the tails of dimensionful observables. In  particular, the presence of EW Sudakov logarithms in \NLOtwo can reduce differential predictions by up to $10\%$. In addition, we have found that for observables affected by large NLO QCD corrections due to real radiation, the \NLOthree contribution can be enhanced. This can lead to an accidental cancellation with \NLOtwo since the origin of both contributions is vastly different. Furthermore, the cancellation might be heavily influenced by the event selection and a possible jet veto. Finally, the \NLOprd prediction is able to mimic the complete NLO calculation for hadronic observables, while it underestimates the complete prediction by up to $1\%-2\%$ for non-hadronic ones. Nevertheless, such small effects are well within the theoretical uncertainties due to scale dependence. Thus, the \NLOprd theoretical prediction can be safely used to model various differential cross-section distributions in phenomenological studies at the LHC for the $pp\to t\bar{t}\gamma+X$ process in the di-lepton top-quark decay channel taking into account the current theoretical precision for this process.

%
\section{Top-quark pair production with two isolated photons}
\label{sec:ttaa}
%

%
\subsection{Integrated fiducial cross sections}
\label{sec:ttaa-int}
%

Next we continue with the $pp\to t\bar{t}\gamma\gamma$ process and start with the discussion of the integrated fiducial cross section, again focusing our attention on the di-lepton top-quark decay channel. In particular, we are interested in any differences between this process and $pp\to t\bar{t}\gamma$.
\begin{table*}[t!]
    \centering
    \renewcommand{\arraystretch}{1.2}
    \begin{tabular}{ll@{\hskip 10mm}l@{\hskip 10mm}l@{\hskip 10mm}}
        \hline
        \noalign{\smallskip}
         &&$\sigma_{i}$ [fb] & Ratio to ${\rm LO}_1$  \\
        \noalign{\smallskip}\midrule[0.5mm]\noalign{\smallskip}
        ${\rm LO}_1$&$\mathcal{O}(\alpha_s^2\alpha^6)$& $ 0.15928(3)^{+31.3\%}_{-22.1\%} $ & $ 1.00 $ \\
        ${\rm LO}_2$&$\mathcal{O}(\alpha_s^1\alpha^7)$& $ 0.0003798(2)^{+25.8\%}_{-19.2\%} $ & $ +0.24\% $ \\
        ${\rm LO}_3$&$\mathcal{O}(\alpha_s^0\alpha^8)$& $ 0.0010991(2)^{+10.6\%}_{-13.1\%} $ & $ +0.69\% $ \\
        \noalign{\smallskip}\hline\noalign{\smallskip}
        ${\rm NLO}_1$&$\mathcal{O}(\alpha_s^3\alpha^6)$& $ +0.0110(2) $ & $ +6.89\% $\\
        ${\rm NLO}_2$&$\mathcal{O}(\alpha_s^2\alpha^7)$& $ -0.00233(2) $ & $ -1.46\% $\\
        ${\rm NLO}_3$&$\mathcal{O}(\alpha_s^1\alpha^8)$& $ +0.000619(1) $ & $ +0.39\% $\\
        ${\rm NLO}_4$&$\mathcal{O}(\alpha_s^0\alpha^9)$& $ -0.0000166(2) $ & $ -0.01\% $\\
        \noalign{\smallskip}\hline\noalign{\smallskip}
        ${\rm LO}$&& $ 0.16076(3)^{+30.9\%}_{-21.9\%} $ & $ 1.0093 $ \\
        ${\rm NLO}_{\rm QCD}$&& $ 0.1703(2)^{+1.9\%}_{-6.2\%} $ & $ 1.0690 $ \\
        ${\rm NLO}_{\rm prd}$&& $ 0.1694(2)^{+1.7\%}_{-5.9\%} $ & $ 1.0637 $ \\
        ${\rm NLO}$&& $ 0.1700(2)^{+1.8\%}_{-6.0\%} $ & $ 1.0674 $ \\
        \noalign{\smallskip}\hline\noalign{\smallskip}
    \end{tabular}
    \caption{\label{tab:ttaa} \it Same as Table \ref{tab:tta} but for the $pp\to t\bar{t}\gamma\gamma +X$ 
    process in the di-lepton top-quark decay channel.}
\end{table*}
In Table \ref{tab:ttaa} we present the integrated fiducial cross section for \LOfull, \NLOfull, \NLOqcd and \NLOprd as well as the corresponding scale uncertainties. We also present the numerical results for all individual ${\rm LO}_i$ and ${\rm NLO}_i$ contributions. We find again that all subleading LO contributions amount to less than $1\%$ of \LOone and are therefore negligible when comparing to the LO scale uncertainties. The \NLOone contribution is similar in size with $6.9\%$ compared to $6.2\%$ for $pp\to t\bar{t}\gamma$. As expected, this contribution yields the largest higher-order corrections at the NLO level. The biggest difference between the two processes is found for \NLOtwo. Indeed, the \NLOtwo contribution increases from $-0.3\%$ to $-1.5\%$, which is consistent with the findings in the literature for the process at hand but with stable top quarks \cite{Pagani:2021iwa}. The \NLOthree contribution remains at the level of $0.4\%$. Finally, \NLOfour contributes at the level of $0.01\%$ only. This contribution is therefore smaller than the MC integration error and phenomenologically completely irrelevant. Due to the increase in size of \NLOtwo, there are larger cancellations between the individual contributions, so that the difference between the \NLOqcd and \NLOfull predictions decreases and both results agree within their coresponding MC errors. As for the $pp\to t\bar{t}\gamma$ process, also in this case the scale uncertainties at the integrated fiducial cross-section level are barely affected by the subleading contributions. Including all subleading LO contributions and subleading NLO corrections to the production of $pp\to t\bar{t}\gamma\gamma$ in \NLOprd leads to a decrease of the \NLOqcd result by about $-0.5\%$. Similar to the previous process, we again have cancellations between the subleading LO contributions ($0.9\%$) and the subleading NLO corrections ($-1.4\%$). The differences between \NLOfull and \NLOprd are reduced to $0.4\%$ compared to $0.9\%$ for the $pp\to t\bar{t}\gamma$ process, implying that subleading NLO corrections involving radiative top-quark decays are less important, at least at the integrated fiducial cross-section level.

%
\subsection{Differential fiducial cross sections}
\label{sec:ttaa-diff}
%

\begin{figure}[t!]
    \begin{center}
	\includegraphics[width=0.49\textwidth]{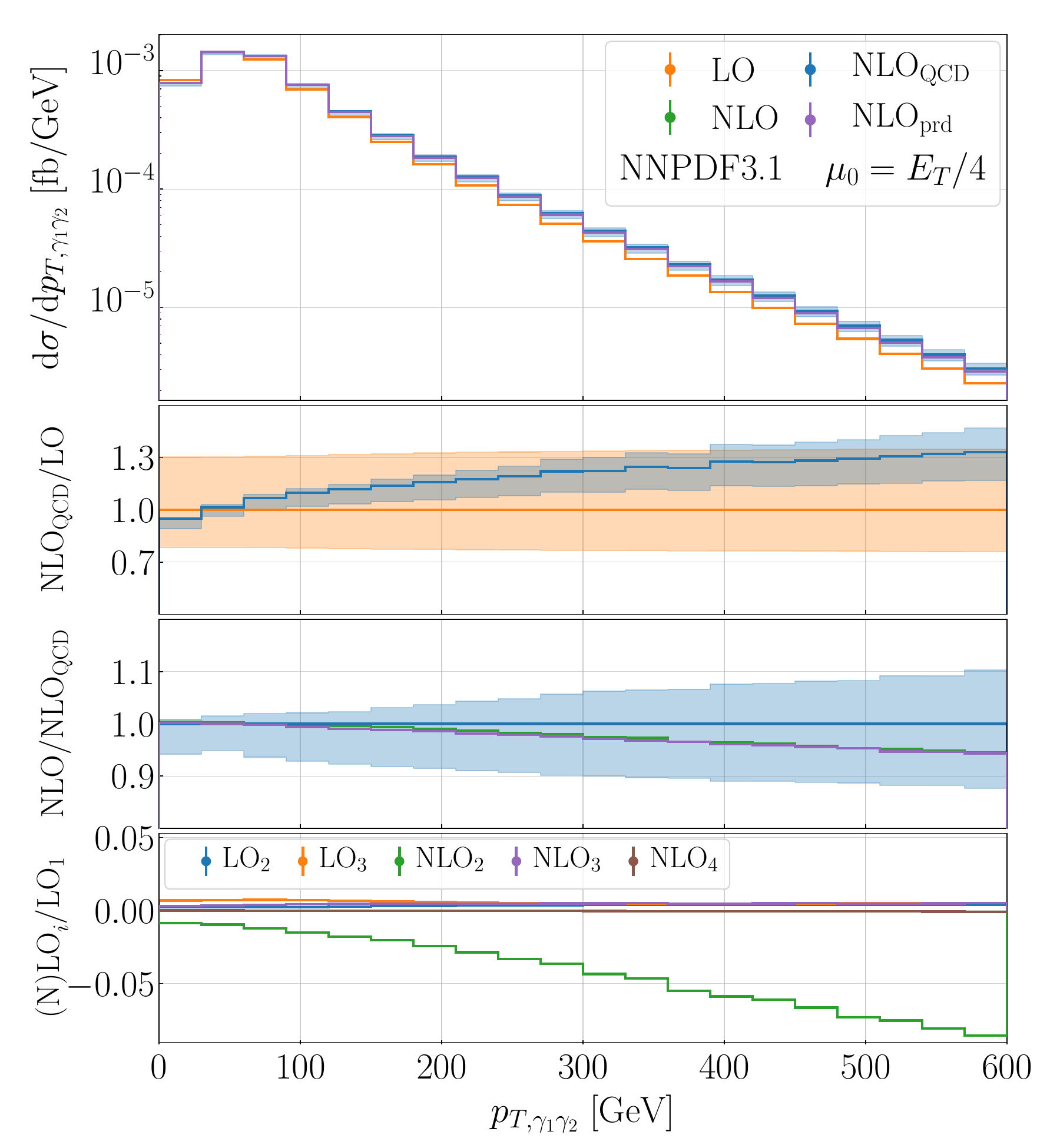}
	\includegraphics[width=0.49\textwidth]{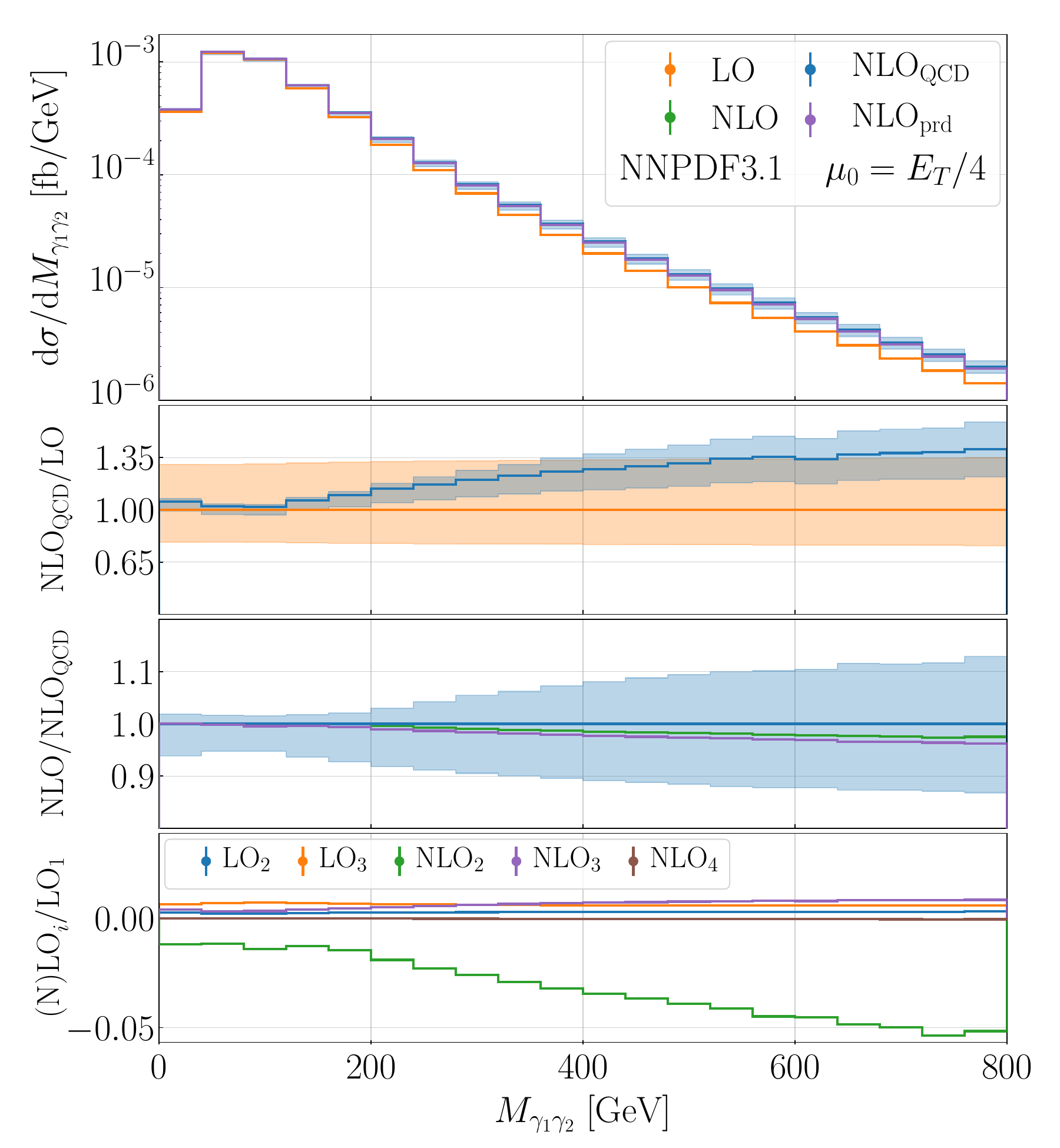}
	\includegraphics[width=0.49\textwidth]{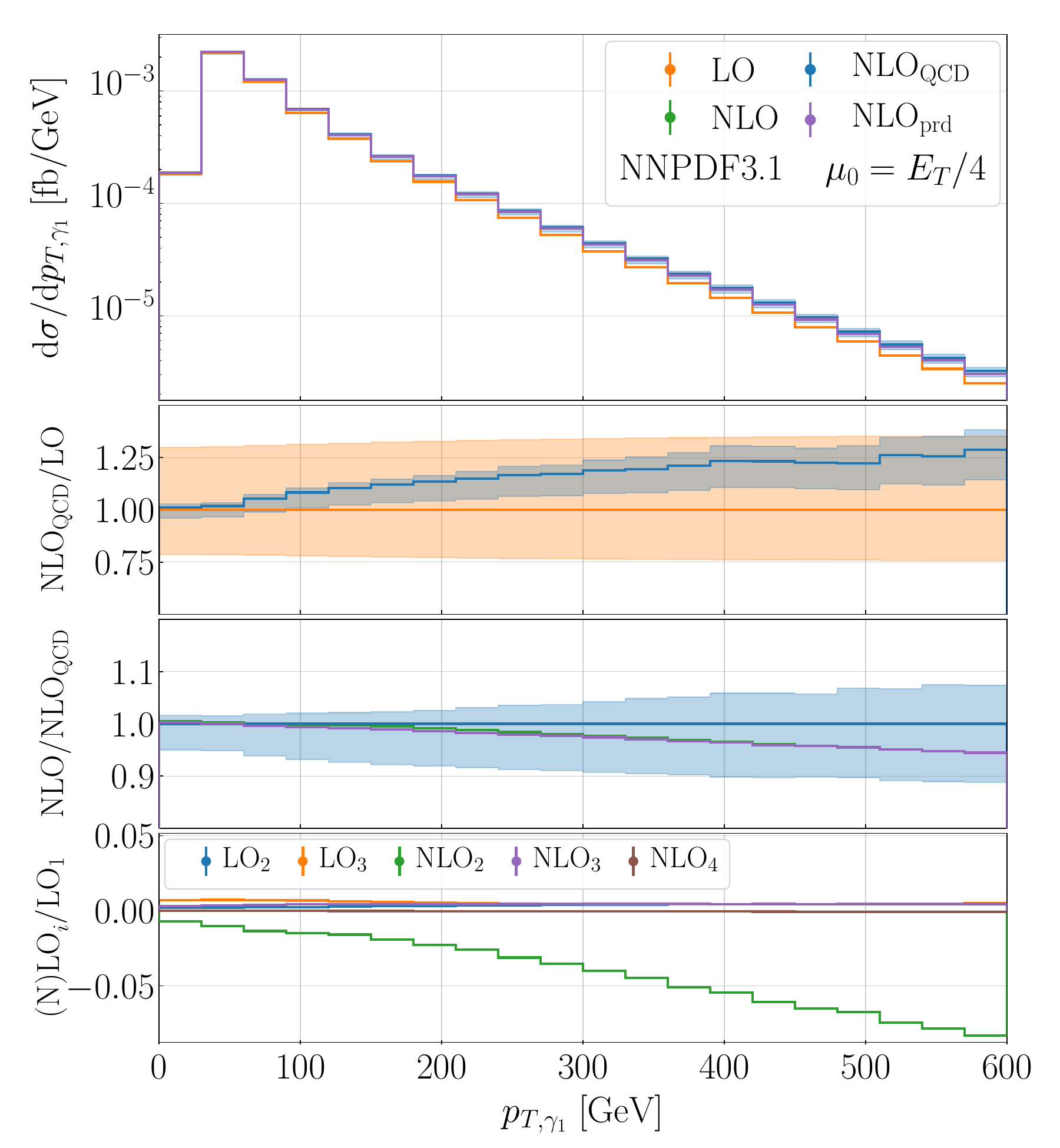}
	\includegraphics[width=0.49\textwidth]{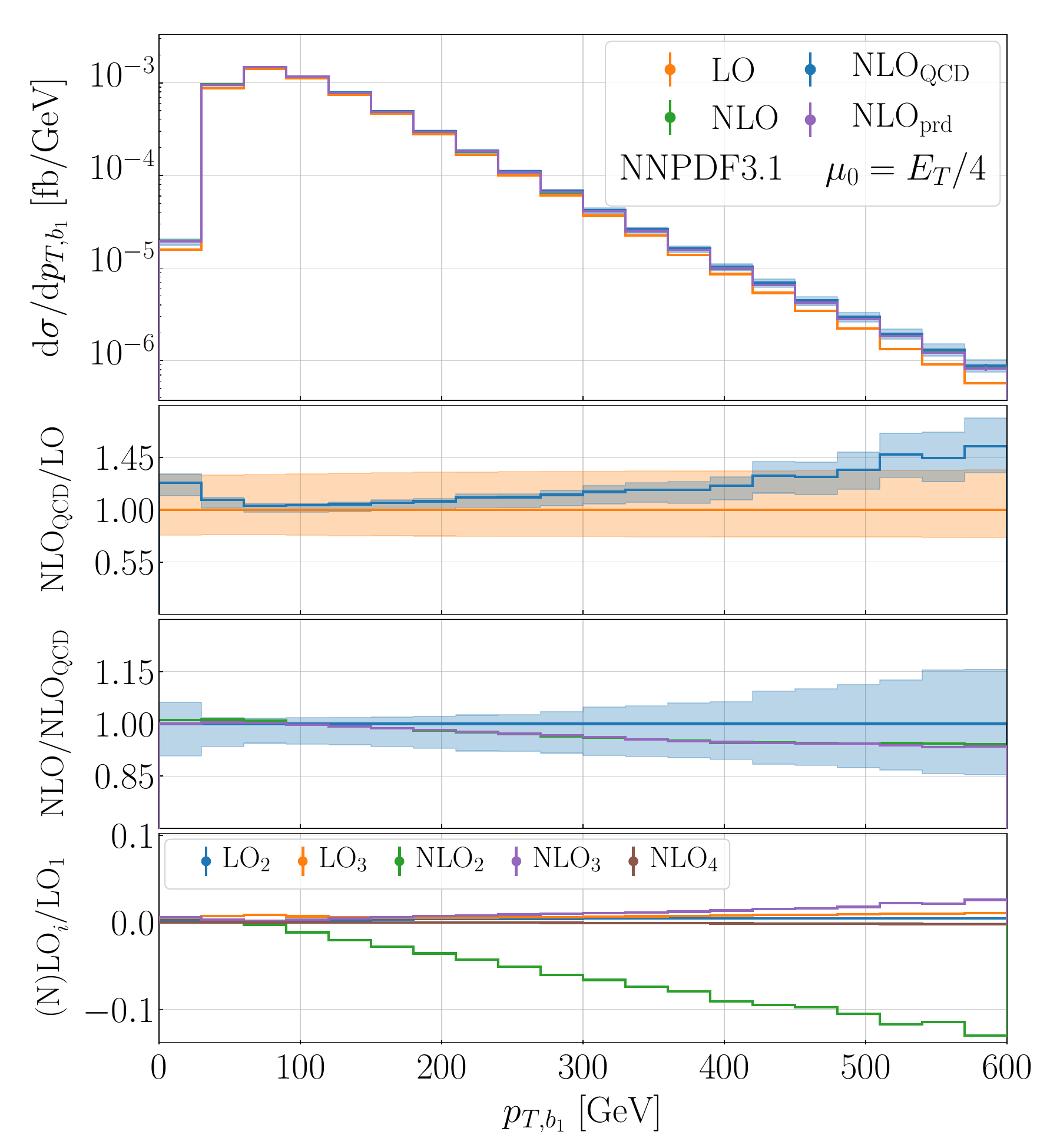}
    \end{center}
    \caption{\label{fig-ttaa:nlo1} \it Same as Figure \ref{fig-tta:nlo1} but for the $pp\to t\bar{t}\gamma\gamma +X$ process in the di-lepton top-quark decay channel and for the observables $p_{T,\gamma_1\gamma_2}$, $M_{\gamma_1\gamma_2}$, $p_{T,\gamma_1}$ and $p_{T,b_1}$.}
\end{figure}

Similarly to $t\bar{t}\gamma$ production, also for the $pp\to t\bar{t}\gamma\gamma+X$ process we are investigating the impact of all subleading contributions and subleading higher-order corrections on various differential cross-section distributions. In Figure \ref{fig-ttaa:nlo1} we show the observables $p_{T,\gamma_1\gamma_2}$, $M_{\gamma_1\gamma_2}$, $p_{T,\gamma_1}$ and $p_{T,b_1}$. As in the case of $pp\to t\bar{t}\gamma+X$,  here we again present in the upper panels the absolute predictions for \LOfull, \NLOfull, \NLOqcd and \NLOprd together with the \NLOqcd uncertainty bands resulting from scale variations. The middle panels show the ratio of  the \NLOqcd results to \LOfull along with their relative scale uncertainties as well as the ratio of the \NLOfull and \NLOprd results to \NLOqcd including the \NLOqcd scale uncertainties. Lastly, the lower panels show the size of all subleading contributions compared to the \LOone result. 
\begin{figure}[t!]
    \begin{center}
	\includegraphics[width=0.49\textwidth]{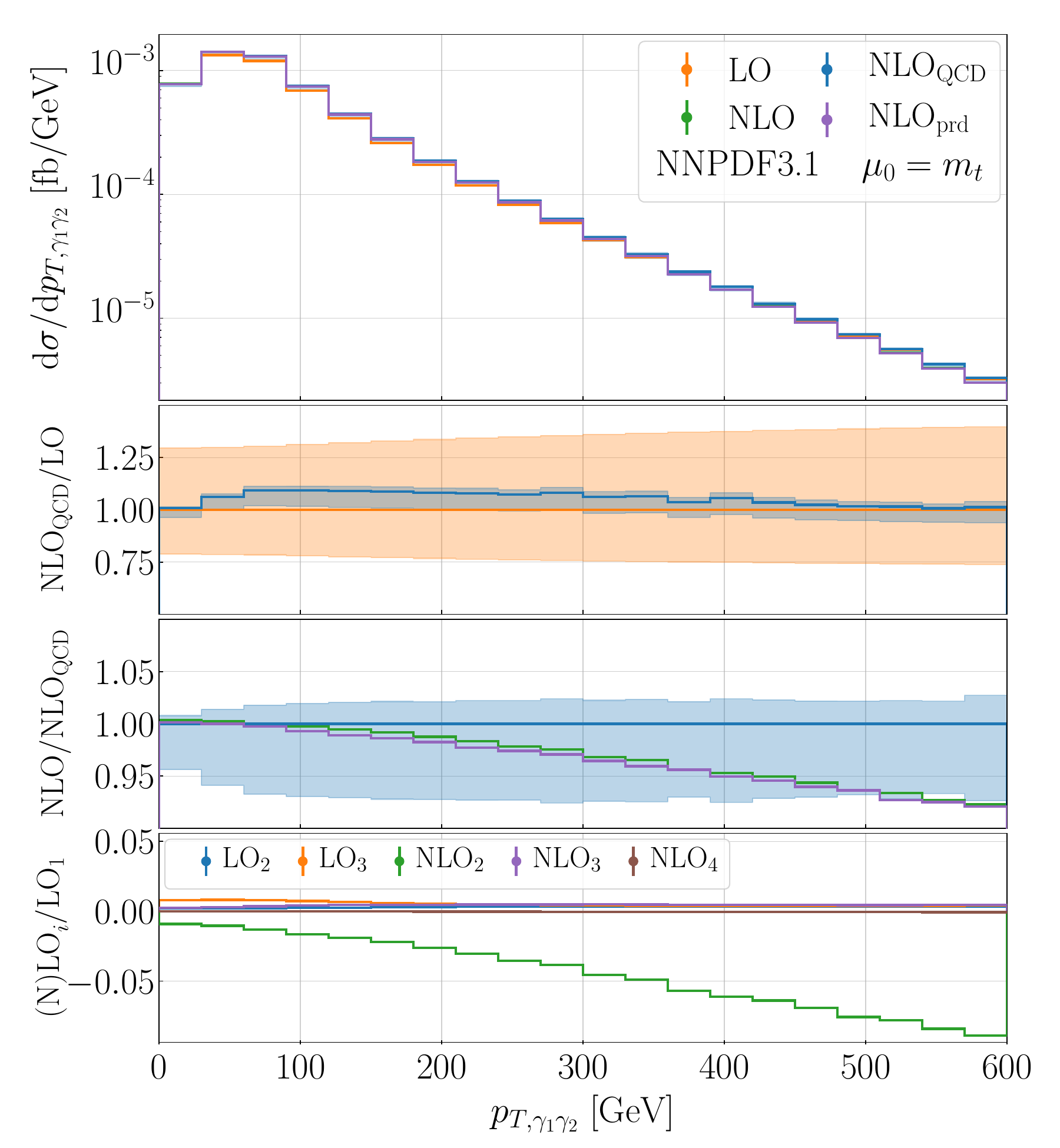}
	\includegraphics[width=0.49\textwidth]{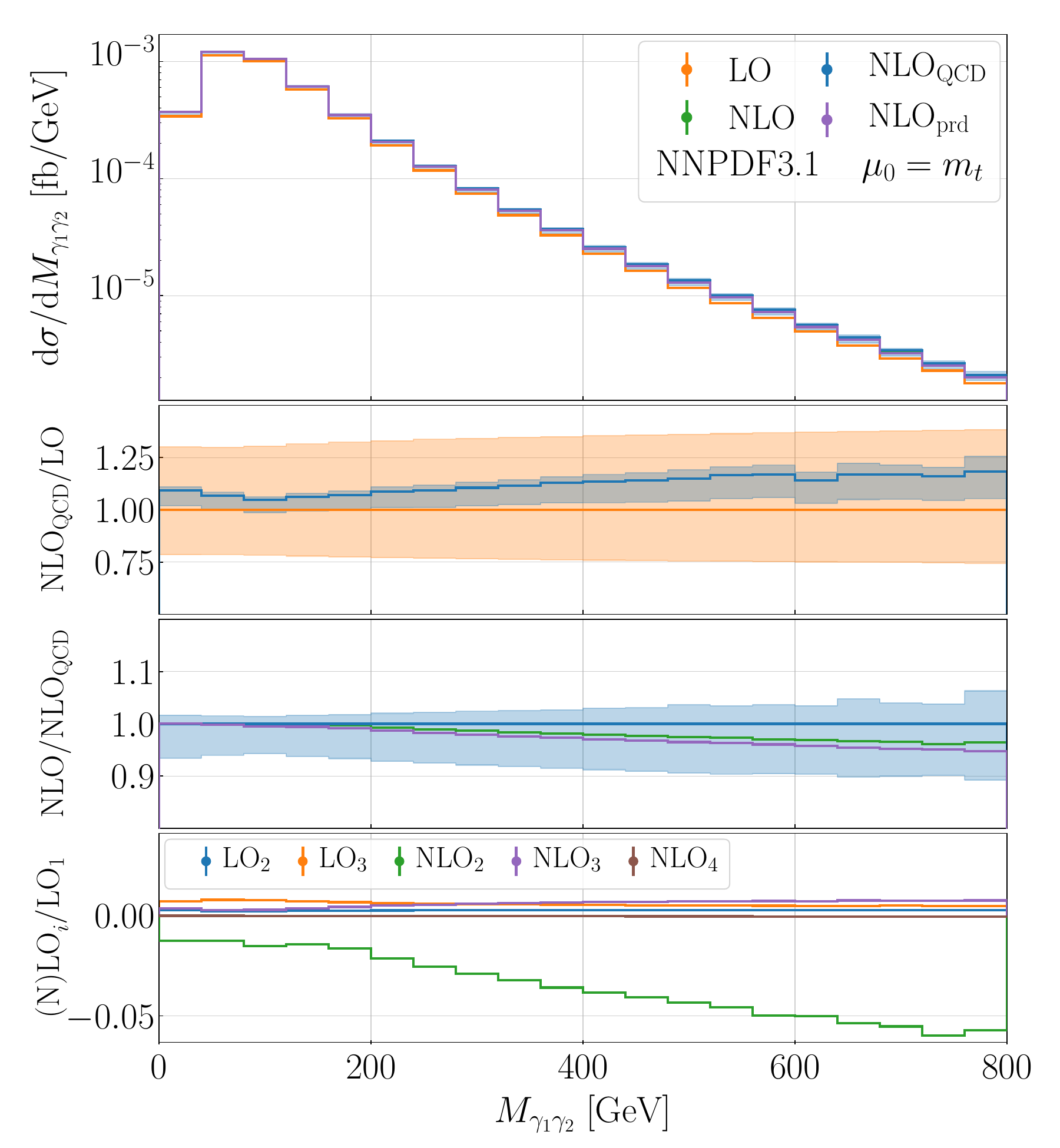}
	\includegraphics[width=0.49\textwidth]{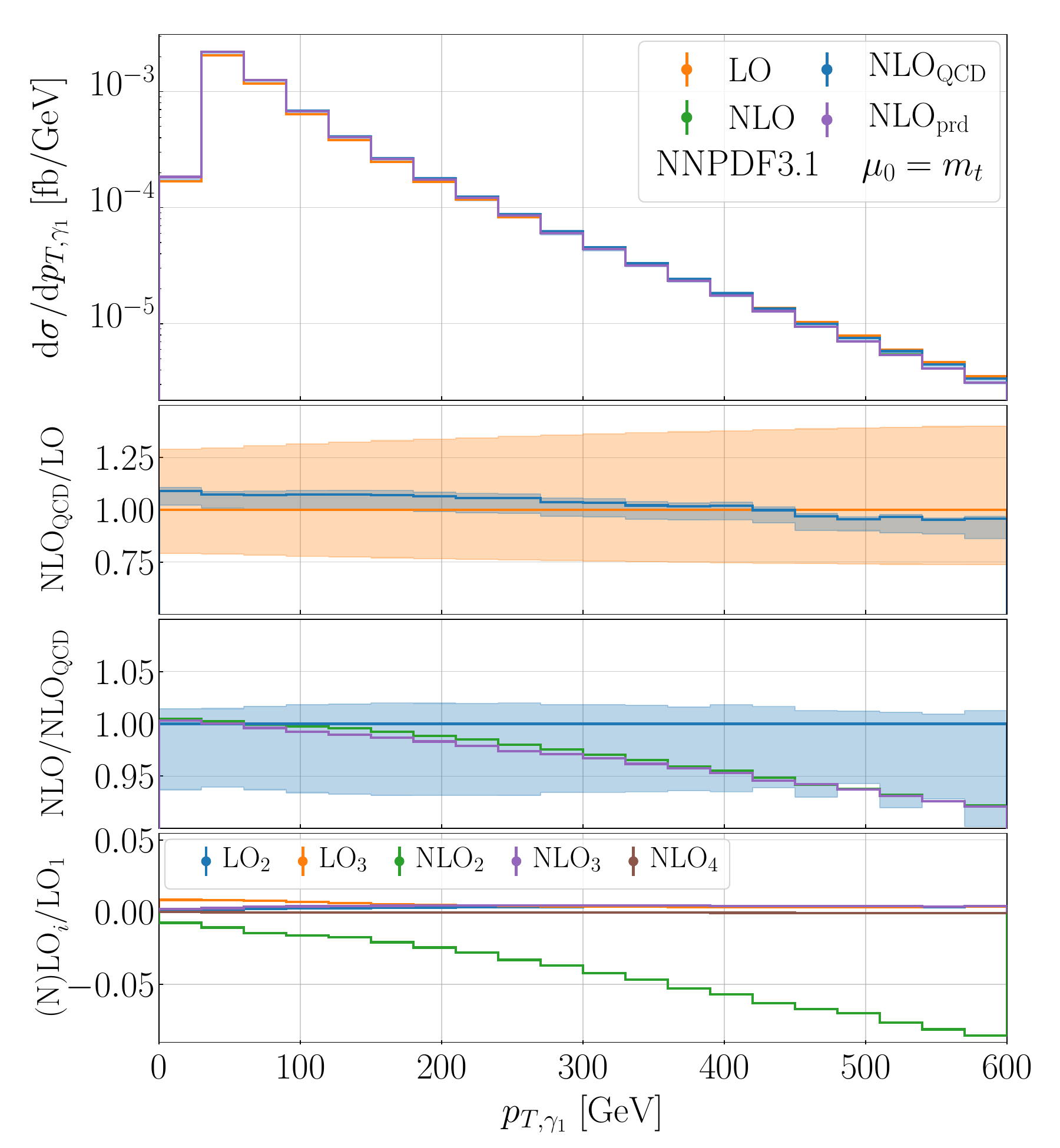}
    \end{center}
    \caption{\label{fig-ttaa:nlo1_mt} \it Same as Figure \ref{fig-tta:nlo1} but for the $pp\to t\bar{t}\gamma\gamma +X$ process in the di-lepton top-quark decay channel and for the observables $p_{T,\gamma_1\gamma_2}$, $M_{\gamma_1\gamma_2}$ and $p_{T,\gamma_1}$.  Results are shown for $\mu_R=\mu_F=\mu_0=m_t$.}
\end{figure}

First, we focus on $p_{T,\gamma_1\gamma_2}$ and $M_{\gamma_1\gamma_2}$. Their importance stems from the fact that both differential cross-section distributions can be seen as the direct and irreducible background to the kinematics of the Higgs boson in $pp\to t\bar{t}H$ production with the $H\to \gamma\gamma$ decay.  For both observables we find that \NLOone is the dominant source of NLO corrections, leading to an increase of \LOone of up to $30\%-40\%$ in the tails. The subleading LO and NLO contributions are dominated by \NLOtwo, due to EW Sudakov logarithms, which amounts to $-8\%$ for $p_{T,\gamma_1\gamma_2}$ and $-5\%$ for $M_{\gamma_1\gamma_2}$ with respect to \LOone. This results in a decrease of the full calculation by about $5\%$ and $3\%$, respectively. At the same time, in these phase-space regions, the \NLOfull scale uncertainties are larger by a factor of $2$ for $p_{T,\gamma_1\gamma_2}$ and $4$ for $M_{\gamma_1\gamma_2}$. The \NLOprd prediction fully recovers the complete calculation for $p_{T,\gamma_1\gamma_2}$ and differences of only about $1\%$ can be found for $M_{\gamma_1\gamma_2}$. Thus, \NLOprd is more than sufficient to  properly describe the shape of the two observables in the presence of subleading LO contributions and subleading NLO corrections.

We then turn to the $p_{T,\gamma_1}$ differential cross-section distribution, where NLO QCD corrections up to $25\%-30\%$ can be found. The subleading contributions reduce the \NLOfull calculation by about $5\%-6\%$ and are therefore slightly larger than those found for the $pp\to t\bar{t}\gamma$ process. Indeed, in the latter case the difference is about $4\%$.  This small rise is due to the increase in the \NLOtwo contribution that we have already observed at the integrated cross-section level. In the case of $p_{T,b_1}$ the difference between \NLOqcd and \NLOfull is somewhat enlarged in the tails from $4\%$ for $pp\to t\bar{t}\gamma$ to $6\%$ for the $pp\to t\bar{t}\gamma\gamma$ process. This can again be attributed to the increase of the \NLOtwo corrections from $-10\%$ to $-13\%$, while the size of \NLOthree is slight reduced to $2\%-3\%$ from to $3\%-4\%$ for $pp\to t\bar{t}\gamma$. In general, the presence of a second photon already at LO affects the kinematics of the $t\bar{t}$ system in such a way that the large NLO QCD corrections in \NLOone are reduced for several observables such as $p_{T,b_1}$ or $p_{T,b_1b_2}$ compared to the $pp\to t\bar{t}\gamma$ process. In particular, for $p_{T,b_1}$ we find a reduction from $95\%$ to $55\%$. It follows directly that the enhancement of \NLOthree in such observables is also reduced. Therefore, the accidental cancellations between \NLOtwo and \NLOthree, which occurred in the $pp\to t\bar{t}\gamma$ case, although still present, are substantially reduced.

For comparison purposes and similar to what we have done for photonic observables in case of the $pp\to t\bar{t}\gamma$ process, also here we show differential cross-section distributions with the alternative scale choice. In detail, in Figure \ref{fig-ttaa:nlo1_mt} we display   $p_{T,\gamma_1\gamma_2}$, $M_{\gamma_1\gamma_2}$ and $p_{T,\gamma_1}$ for $\mu_R=\mu_F=m_t$. For $p_{T,\gamma_1\gamma_2}$ and $M_{\gamma_1\gamma_2}$, we again find that this scale choice leads to a reduction in the size of the NLO QCD corrections in \NLOone from about $30\%$ to $10\%-12\%$. In addition, the scale uncertainties are changed from $12\%-13\%$ to $8\%-10\%$. Thus, the fixed scale setting not only decreases higher-order corrections in \NLOone, but also provides improved scale uncertainties. The reduction of higher-order effects in \NLOone has also a direct impact on the significance of the subleading NLO corrections. In particular, because the relative size of the subleading \NLOtwo corrections with respect to \LOone does not change for $\mu_0=m_t$, the overall \NLOtwo contribution becomes larger with respect to  the complete \NLOfull result. Indeed, we observe the rise of \NLOtwo from $5\%$ to $8\%$ as well as from $3\%$ to $4\%$ for $p_{T,\gamma_1\gamma_2}$ and $M_{\gamma_1\gamma_2}$, respectively. Furthermore, for $p_{T,\gamma_1\gamma_2}$ the subleading NLO corrections become as large as the corresponding scale uncertainties of \NLOqcd, which are also equal in size to those of the complete \NLOfull result. The same can be observed for the third differential cross-section distribution, namely for $p_{T,\gamma_1}$. In this case the \NLOone contribution is reduced in the tails from $20\%$ for $\mu_0=E_T/4$ to $5\%$ for $\mu_0=m_t$. This increases the importance of the \NLOtwo contribution again and leads to a reduction of \NLOqcd by about $8\%$ compared to $5\%-6\%$ for $\mu_0=E_T/4$. Consequently, the subleading contributions are of similar size as the scale uncertainties of the \NLOqcd result, and are at the level of $10\%$.  Lastly, for all three photonic observables, the complete \NLOfull predictions for the two scale settings differ by at most $4\%$ in the tails and are therefore within the respective scale uncertainties.

%
\section{Summary}
\label{sec:sum}
%

We have presented the first calculation of the complete set of NLO corrections to $pp\to t\bar{t}\gamma+X$ and $pp\to t\bar{t}\gamma\gamma+X$ including top-quark decays at the LHC with $\sqrt{s}=13$ TeV. In order to study the di-lepton top-quark decay channel we have employed the  Narrow Width Approximation  that works in the $\Gamma/m\to 0$ limit and preserves  spin correlations. In this approximation all contributions without two resonant top quarks and $W$ gauge bosons are simply neglected. In our calculations we have included the dominant LO contribution at ${\cal O}(\alpha_s^2\alpha^5)$ for $pp \to t\bar{t}\gamma +X$ and ${\cal O}(\alpha_s^2\alpha^6)$ for $pp\to t\bar{t}\gamma\gamma+X$ as well as the corresponding NLO QCD corrections. Furthermore, all subleading LO contributions and all remaining NLO corrections are also taken into account. In addition, NLO corrections, as well as photon radiation are consistently included in the production phase as well as in all decay stages of the process. Even if we have considerd the di-lepton decay channel of the top quark, the extension to hadronic decays of the $W$ boson can be straightforwardly incorporated into our framework.

On the technical side, to perform these calculations, we have extended the Nagy-Soper subtraction scheme as implemented in the \textsc{Helac-Dipoles} Monte Carlo program for calculations with QED-like singularities and modified the structure of the program to allow the simultaneous calculation of all contributions at different orders in $\alpha_s$ and $\alpha$.

The main findings of this paper apply to both  processes and can be summarised as follows. At the integrated fiducial cross-section level, as well as for all the angular distributions we examined, the subleading LO contributions and subleading NLO corrections are negligibly small with respect to the \NLOfull scale uncertainties of the complete result. On the other hand, the dominant \NLOone contribution is essential for precise predictions, as it leads to a reduction in scale uncertainties from about $30\%$ to $6\%$.  Furthermore, it also introduces a change in the normalisation as well as in the shape of various differential cross-section distributions. In general, the inclusion of the subleading contributions has been found to be essential only in the tails of dimensionful observables due to one-loop EW Sudakov logarithms. Indeed, the presence of the EW Sudakov logarithms in \NLOtwo  leads to a reduction in the tails of up to $10\%$ compared  to the \NLOqcd result. Moreover, the \NLOtwo contribution can be as large as the \NLOqcd scale uncertainties, potentially affecting the comparison between theoretical predictions and experimental measurements. In addition, for certain observables affected by large higher-order real-emission QCD corrections, e.g. for $p_{T,b_1b_2}$, the \NLOthree contribution can be enhanced from a few percent up to $10\%$ with respect to \LOone.  As the origins of  the subleading \NLOtwo and \NLOthree contributions are very different and there are random cancellations between them, they should always be considered together. We note here, that both the accidental cancellations between \NLOtwo and \NLOthree as well as the size of real-emission QCD corrections depend on the exact event selection and can be substantially affected, for example, by applying a veto on an additional jet. Finally, the subleading LO contributions and the \NLOfour contribution are negligibly small at the integrated and differential fiducial cross-section level with respect to the NLO scale uncertainties. 

An important finding of the paper is that the \NLOprd  approximation models the complete \NLOfull result very well. Indeed, differences of up to $2\%$ only are found for some leptonic and/or photonic observables, but these are negligible compared to the NLO scale uncertainties of the complete result. We remind the readers that in \NLOprd, all LO contributions as well as \NLOone are fully included in both the $t\bar{t}\gamma(\gamma)$ production and decays of the top-quark pair. However, the subleading NLO corrections (\NLOtwo, \NLOthree, \NLOfour) are only included in the production stage of the $t\bar{t}\gamma(\gamma)$ process. The same applies to photon radiation for these three NLO contributions. The \NLOprd approximation not only provides a great simplification of higher-order calculations, especially when it comes to the real emission corrections, but will ultimately also be of great benefit when matching \NLOprd predictions to parton shower programs. Indeed,  radiation from unstable paricles and their decay products  can lead to severe unphysical distortions of the intermediate resonances, if not properly treated. This problem was first pointed out in the context of NLO QCD plus parton shower simulations for top-quark pair production and decay \cite{Campbell:2014kua,Jezo:2015aia}, and was solved in the context of the \textsc{Powheg Box} framework by means of the so-called resonance-aware matching \cite{Jezo:2016ujg,Jezo:2023rht}. However, a method for consistently combining the radiation emitted at the complete-NLO level with a QCD+QED parton shower in the presence of non-trivial resonances is not yet available in the literature. Therefore, the simplification in \NLOprd of not including subleading corrections in the decays, which would allow the matching of subleading contributions with standard procedures like \textsc{Powheg} \cite{Nason:2004rx,Frixione:2007vw,Alioli:2010xd} and/or \textsc{MC@NLO} \cite{Frixione:2002ik,Frixione:2003ei} (once they are available for the complete NLO case), is highly desirable and of great importance.

\acknowledgments{

We thank Ansgar Denner and Mathieu Pellen for providing us with the results for the NLO EW  corrections to the $pp \to e^+\nu_e \mu^- \bar{\nu}_\mu b\bar{b} +X$  process that helped to cross check our implementation of the EW corrections in  the \textsc{Helac-Dipoles} framework.

This work was supported by the Deutsche Forschungsgemeinschaft (DFG) under grant 396021762 $-$ TRR 257: {\it P3H - Particle Physics Phenomenology after the Higgs Discovery}. Support by a grant of the Bundesministerium f\"ur Bildung und Forschung (BMBF) is additionally acknowledged.

The authors gratefully acknowledge the computing time provided to them at the NHR Center
NHR4CES at RWTH Aachen University (project number {\tt p0020216}). This is funded by the Federal
Ministry of Education and Research, and the state governments participating on the basis of the
resolutions of the GWK for national high performance computing at universities.}


\bibliographystyle{JHEP}


\providecommand{\href}[2]{#2}\begingroup\raggedright\endgroup

\end{document}